\documentclass[a4paper, 12pt, twoside]{article}
	\usepackage[utf8]{inputenc}
	\usepackage[T1]{fontenc}
	\usepackage{lmodern}
    \usepackage[french,english]{babel}
    \usepackage{csquotes}
    \usepackage{listings}
    \usepackage[left=2.5cm,right=2.5cm,top=3cm,bottom=3cm]{geometry}
    \usepackage[noblocks]{authblk}

	\usepackage{amsmath}
	\usepackage{amssymb}
	\usepackage{bm}
	\usepackage{siunitx}
	\usepackage{braket}
	\usepackage{cancel}
	\usepackage{subcaption}

	\usepackage{graphicx}
	\usepackage[x11names,table]{xcolor}
	\usepackage{pgfplots}
	\pgfplotsset{width=7cm,compat=1.13}
	\usepackage{url}
	\usepackage{doi}
	\usepackage{hyperref}
	  \hypersetup{colorlinks,%
      linkcolor=Blue3,%
      urlcolor=Blue3,%
      citecolor=Red4,%
      filecolor=Green4}
     \usepackage[
      backend=biber,
      style=numeric,
      giveninits=true,
      natbib=true,
      sorting=none,
      url=false, 
      doi=true,
      eprint=true
      ]{biblatex}

\begin{document}

\title{Double scaling limit of the prismatic tensor model}

%{\small
\author[1]{T.~Krajewski} 
\author[2]{T.~Muller \thanks{corresponding author}}
\author[2,3]{A.~Tanasa}

\affil[1]{Aix Marseille Univ, Université de Toulon, CNRS, CPT, Marseille, France, EU}
\affil[2]{Univ. Bordeaux, LaBRI CNRS UMR 5800, Talence, France, EU}
\affil[3]{DFT, H. Hulubei Nat. Inst. Phys. Nucl. Engineering, Magurele,  Magurele, Romania, EU}

%}

\maketitle

\abstract

In S. Giombi, I. Klebanov, F. Popov, S. Prakash and G. Tarnopolsky, {\it Phys. Rev.} {\bf D}  98 (2018) 10, 105005, 
a prismatic tensor model was introduced.
We study here the diagrammatics and the double scaling limit of this model, using the intermediate field method. 
We explicitly exhibit the next-to-leading order Feynman graphs in the $1/N$ expansion. 
Using appropriate combinatorial tools, we further study the general term of the $1/N$ expansion 
and we 
compute the $2-$point function in the double scaling limit.
We find that 
the double scaling limit mechanism implemented here is similar to the one implemented for various quartic tensor models.

\iffalse
\begin{titlepage}
    \begin{center}
        \vspace*{1cm}
            
        \Huge

        \vspace{0.5cm}
     
        \vspace{0.5cm}
        \Large
         
        T. Krajewski, T. Muller and A. Tanasa

        \Large
       
    \end{center}
\end{titlepage}
\fi

\tableofcontents
\section{Introduction}

Tensor models are zero-dimensional quantum field theoretical models where the fields are tensors of arbitrary rank 
%and whose components are randomly distributed . 
 (see the books 
 \cite{gurau_2017} 
 \cite{book_Tanasa}
 or the review articles
 \cite{Tanasa_2016},
\cite{Tanasa:2012pm}, 
 \cite{https://doi.org/10.48550/arxiv.1907.03531}, 
 \cite{Gurau-aihpd}
or \cite{TASI}).

Introduced as a generalization 
in dimension three and higher
of 
the celebrated
random matrix models (see \cite{Zuber} and the reviews \cite{Francesco_1995,Francesco_2004}), tensor models are currently studied  in the context of mathematical physics, combinatorics
and various other domains of mathematics. 
Thus, rank $d$ tensor models are naturally linked to $d$ dimensional random geometries, as the Feynman graphs present in their perturbative expansion are dual to random triangulations.
Moreover, such models were proven by Witten \cite{https://doi.org/10.48550/arxiv.1610.09758} 
and then by Klebanov and Tarnopolsky
\cite{Klebanov}
to be related to the holographic Sachdev-Ye-Kitaev model \cite{Kitaev}.
%that was proposed as a toy model of holography. 
%Random matrix models, generalized by the tensor models, were also proven to be related to 2D quantum gravity \cite{DiFrancesco:1993cyw} \cite{DiFrancesco:2004qj} and many other subjects in physics and mathematics \cite{Zuber}. 

Initially proposed in the nineties (see, for example, \cite{doi:10.1142/S0217732391001184}), tensor models 
%did not draw a particular amount of interest because of the 
lacked the implementation of the large $N$ mechanism ($N$ being in this context the size of the tensor).
This changed in 2011, with the proposition of the so-called colored tensor model by Gurau \cite{Gurau_2011}, which he proved to have a well-defined large $N$ expansion \cite{Gurau_2011_2}. 
Shortly after, a less restricted model (from the point of view of the class of Feynman graphs allowed by the perturbative expansion),  the multi-orientable  model \cite{MO} was proposed and proven to have a well-defined large $N$ expansion \cite{Dartois_2013}.  
%These models are based on invariances under the action of different symmetry groups. The one considered in this internship, called the $O(N)^3$ invariant-tensor model, 

%An even more general tensor model, 
%possessing an action invariant under the group 
An $O(N)^3$-invariant tensor model, was proposed in \cite{Carrozza_2016} (see also
\cite{Bonzom_2022,Nador2019} for various diagrammatic developments). 
We refer to this type of model as single field tetrahedric model, since one has only one tensor field in the action, and the interaction is tetrahedric (its index
structure has the topology of a tetrahedron).

Let us also mention that the model proposed in \cite{Carrozza_2016} was later related by Klebanov and Tarnopolsky 
in \cite{Klebanov}
to the celebrated holographic SYK model.

%Recently, t
The case of sextic interactions for this type of $O(N)^3$ invariant-tensor model was studied by Giombi et al. in \cite{Giombi_2018}. 
This model was called prismatic since its large $N$ limit is dominated by a positive-definite operator whose index
structure has the topology of a prism. In this paper, we will study this prismatic model.

Let us recall that the large $N$ expansion of tensor models is controlled, as usually in tensor models, by a parameter $\omega$ called the degree. 
%Contrary to matrix models where this mechanism is controlled by the genus $g$ of the graphs, the degree is not a topological invariant. Moreover, in the case of tensor models, t
The large $N$ expansion is dominated by graphs of vanishing degree.
%, that are often called leading order (LO) graphs. 
%Both the expression of the degree and the structure of the graphs depend on the interaction considered in the model \cite{Carrozza_2016}. 
%For the prismatic model 
In \cite{Prakash_2020}, it was proven that 
the leading order (LO) graphs 
in the large $N$ expansion of the prismatic model 
are not the usual melonic graphs one is used to from tensor model literature
 (see again the books 
 \cite{gurau_2017} 
 \cite{book_Tanasa}
 or the review articles
 \cite{Tanasa_2016},
\cite{Tanasa:2012pm}, 
 \cite{https://doi.org/10.48550/arxiv.1907.03531}
or \cite{TASI}).
The role of the fundamental melon is played by a triple tadpole graph. Moreover, one now has two types of melonic insertions, one at the level of the propagator (as usually in tensor model literature), and another one at the level of the vertex, which is a new type of melonic move with respect to what it is known in the tensor model literature.

In this paper we give a different proof of the LO graph identification of \cite{Prakash_2020}. Our proof relies on an extensive use of the intermediate field method, which expresses the prismatic model as a model with quartic tetrahedric interactions, but with two distinct tensor field (the initial tensor field and an intermediate field).

Moreover, the intermediate field method used in this papers allows us to identify the next-to-leading (NLO) graphs in the large $N$ expansion.

The main result of this paper is then the implementation of the celebrated double scaling limit mechanism for the prismatic tensor model.
%A second mechanism, richer than the large $N$ limit, can be studied. Called double scaling limit (DSL), it 
%consists of taking both $N$ to infinity and the coupling constant $t$ of the interaction to a critical value $t_c$. This mechanism is controlled by keeping a fixed relation between $N$ and $t$, encoded by the double scaling parameter $\kappa (N,t)$. 
Using a combinatorial scheme decomposition \textit{a la} \cite{Chapuy} allows to identify at any degree a sub-family of graphs, called dominant, whose amplitude is the most divergent when 
%$t \rightarrow t_c$. 
the coupling constant of the model goes to criticality.
%The double scaling parameter is then chosen such that both the LO graphs and the dominant graphs at any degree contribute to the correlation functions.
One then has, in this double scaling limit, contributions of Feynman graphs of any degree, as it was already noticed for 
%The double scaling mechanism was already implemented for different tensor models such as 
the colored \cite{Schaeffer} model,  the multi-orientable model \cite{Gurau_2015} and the single field  tetrahedric model \cite{Bonzom_2022}. 

%The DSL for these models has a tree like behavior. The dominant graphs are in bijection with a specific type of rooted binary trees and 

Note that in this paper we work with the $2-$point function, and not with the free energy when implementing the double scaling mechanism. This comes from the fact that, from a combinatorial point of view, it is easier to work with the $2$-point function, which correspond to what combinatorists call a {\it rooted} graph (and which has no symmetries). 
This is a standard tool in combinatorics and it has already been exported in mathematical physics for the study of quartic tensor models (see again \cite{Bonzom_2022}, \cite{Gurau_2015} and \cite{Schaeffer}). However, the results obtained here for the $2$-point function can be extended, leading to an 
analogous behaviour,
for the free energy (or for $2k$-point function, $k$ here being arbitrary, see \cite{Gurau_2015}, where this was implemented for the multi-orientable tensor model).

Moreover, we prove that
the $2$-point function of the model is summable in the double scaling limit.
%leading to an exact result. Again, 
This behaviour is different with respect to the 
%marks a difference with 
matrix model case.
\medskip

%the 
%diagrammatics of the
%large $N$ expansion (leading order (LO), next-to-leading order (NLO)) of this prismatic model. 
%by S. Giombi et. al. in \cite{Giombi_2018}.}

%In order to do so, we use the intermediate field method, already
%mentioned in  \cite{Giombi_2018}, since
%this method allows to reduce the sextic interaction to a quartic one, 
%%with two tensor fields and since the $O(N)^3-$invariant quartic tensor model has been throughy investigated (see again \cite{Carrozza_2016}
%\cite{Bonzom_2022} or \cite{Nador2019}.
%which is the tetrahedric one.

%The overall aim of this paper is to realize a thorough study 
%of the diagrammatics of the large $N$ expansion (leading order (LO), next-to-leading order (NLO)) of the prismatic model and implement its double scaling limit.

In this paper, our overall strategy is the following:
\begin{enumerate}
%\item  
%generalizing the methods developed for models with single fields quartic interactions to ones with a second field.
\item Apply
the intermediate field method in order to reduce the sextic interaction
of the prismatic model to the quartic, tetrahedric one. As mentioned above, by doing so, one now has two fields in the action, the original tensor field and the intermediate field (also a tensor field). We call this the tetrahedric representation of the model (by contrast to the "original" one, using the prismatic vertex, which we call the prismatic representation of the model).
\item Generalize the results known in the case of the single field tetrahedric model mentioned above to the two-field model obtained 
{\it via} the use of the intermediate field method.
%these methods to the case of the interaction obtained from the intermediate field method.
\item When possible, deduce the results on the original prismatic model from the 
two field model above (thus somehow applying an "inverse" intermediate field method).
\end{enumerate}
Let us mention that deducing some of the results on  the original prismatic model from the 
two field model above does not appear to us to be direct, since some of the combinatorial tools used here are 
natural for quartic interactions but are not transposable in a direct way from the quartic model to the sextic model.
 
\medskip

Note however that even though in \cite{Giombi_2018} the prismatic model is studied in dimension $3-\varepsilon$,
the double scaling limit mechanism we implement in this paper is done for a $0-$dimensional prismatic tensor model. It does not seem clear to us if our results can easily generalize to the dimension $3-\varepsilon$ case.

For the sake of completeness, let us also mention that some sixth order tensor models (but not the prismatic model studied in this paper), as well as other types of tensor models,  have also been studied in \cite{Lionni_2019}, where the authors have
analysed the dominant regime and critical behaviour of their models.

\medskip

The paper is structured in the following way. In the next section,
we give a short review of the double scaling limit mechanism of matrix and quartic tensor models.
In the third section, we recall the definition of the prismatic model and the use of the intermediate field method in this case. In the fourth section, we exhibit the LO graphs, in both the tetrahedric and the prismatic representations of the model. 
%We thus exhibit here the two types of melonic moves found in the prismatic representation, the edge and the vertex one.
In the following section, we study the generating functions of the LO graphs, in both representations of the model.
In the sixth section we introduce appropriate diagrammatic tools in order to study, from a combinatorial point of view, the general term of the 
large $N$ expansion of this model. We use these tools in the following sections, where the NLO Feynman graphs are exhibited. In the eighth section we identify the dominant graphs and compute the $2-$point function is the double scaling limit.
%$G_{2,DS}(t)$.
In the ninth section we obtain, from the $2-$point function in the double scaling limit, the free energy in the double scaling limit.
We end the paper with a concluding section.

%The structure of this report is the following. In Section \ref{Def_mod}, the $O(N)^3$ invariant-tensor model, its perturbation theory, and its graphical interpretation will be first presented. Then, we will restrict the model to a prismatic interaction, explain the intermediate field method, and apply it to the case studied. In Section \ref{large N}, we will investigate the large $N$ expansion of the model. We will start by implementing the $\frac{1}{N}$ expansion and derive the expression of the parameter controlling it. We will then identify the leading order graphs in the expansion and interpret them in the original prismatic model. Finally, we will enumerate the number of leading order graphs with a given number of vertices using Lagrange inversion theorem. In Section \ref{Diagrammatic Analysis}, we will introduce the tools required to study the double scaling limit of the model. We will explain the notion of hybrid line in a graph and their necessity. We will then compute the generating functions of the so called "melon graphs" that lead the $\frac{1}{N}$ expansion as well as deriving their singularities. We will finally define the notion of dipoles, chains, and schemes and explain their utilities in studying graphs that appear in the general term of the $\frac{1}{N}$ expansion. In Section \ref{NLO}, we will identify the next to leading order graphs. 
%Finally, in Section \ref{Finiteness}, we will prove the finiteness of the number of schemes and identify the dominant graphs in the double scaling limit.

\section{Brief review on the double scaling limit}

In this section, we give a short review of the large $N$ limit and the double scaling limit of matrix and quartic tensor models.

For both models, the quantities of interests, such as the partition function, can be expanded in terms of Feynman graphs and rearranged in powers of $\frac{1}{N}$. However, the nature of this expansion differs between the two models.  

On the one hand, matrix models \cite{DiFrancesco:2004qj,DiFrancesco:1993cyw} expand in ribbon graphs. This $\frac 1N$-expansion 
is controlled by the genus $g$ of these Feynman ribbon graphs. The partition function of these models writes:
\begin{equation*}
    Z_M(\lambda)= \sum_{g=0} N^{2-2g} Z_g(\lambda) \text{ ,}
\end{equation*}
where $\lambda$ is the coupling constant of the interaction and $Z_g(\lambda)$ is the contribution of all the genus $g$ graphs.
In the large $N$ limit, only $Z_0(\lambda)$ contributes - planar graphs are dominating the expansion.

On the other hand, rank $D$ quartic tensor models expand in terms of $D+1$ edge-colorable graphs and their $\frac{1}{N}$ expansion is controlled by Gurau's degree $\omega$. The partition function of these models writes:
\begin{equation*}
    Z_T(\lambda) = \sum_{\omega} N^{3-\omega} Z_{\omega}(\lambda)
\end{equation*}
Unlike the genus $g$, the degree is not a topological invariant. Again, in the large $N$ limit only the term $Z_{0}(\lambda)$ contributes, which corresponds to the so-called melonic graphs.
%when considering a tetrahedric interaction.

\bigskip

%The double scaling limit of matrix and tensor models are also quite different. 
Let us now recall the double scaling limit mechanism for matrix and quartic tensor models.
In the case of matrix models, one can prove that the coefficients at each order in the $\frac{1}{N}$ expansion behave as
\begin{equation*}
    Z_g(\lambda) \sim f_g (\lambda - \lambda_c)^{\frac{5}{2}(1-\gamma_{str})g/2} \text{ ,}
\end{equation*}
with $\lambda_c$ a critical value of the coupling constant where all the $Z_g$ diverges and $\gamma_{str}$ a critical exponent of the model.
 By defining a double scaling parameter $\kappa$ as 
\begin{equation*}
    \kappa^{-1} = N (\lambda - \lambda_c)^{2-\gamma_{str}/2} \text{ ,}
\end{equation*}
one can rewrite the expansion as
\begin{equation*}
    Z_M(\lambda)= \sum_{g=0} \kappa^{2g-2} f_g \text{.}
\end{equation*}
The double scaling limit is then obtained by taking $N \rightarrow \infty$ and $\lambda \rightarrow \lambda_c$ while keeping $\kappa$ constant. In this limit, $Z_M$ has contribution of all graphs and not simply the planar ones however the series of $Z_M(\lambda)$ diverges.

For quartic tensor models, the double scaling limit mechanism is obtained in the following way. 
In the tensor model literature,  the double scaling limit mechanism is to our knowledge only implemented for the $2-$point or $2r$-point functions ($r$ beeing an integer).
However, the free energy in the double scaling limit can obtainable from the $2-$point function, and this is explained 
%how at the end of 
in 
Section \ref{Def_mod} below. 

%Second, compared to matrix models only the graphs called "dominant" contribute in the double scaling limit. 
It was found for different models with tetrahdric interaction that the
dominant 
graphs in the double scaling limit are bijection with rooted binary trees and that the $2-$point function had an exact expression \cite{Schaeffer,Bonzom_2022,Gurau_2015}. For example, in the case of the $O(N)^3$-invariant model with tetrahedric interaction one finds
\begin{equation*}
    G_{2}^{DS}= M_c \left( 1+ N^{11/12} \frac{\sqrt{3} \lambda_c^{1/2}}{(1+6 \lambda_c^2)^{1/2}} \frac{1- \sqrt{1-4 \kappa}}{2 \kappa} \right) \text{,}
\end{equation*}
with $\kappa$ the double scaling parameter proportional to
\begin{equation*}
    \kappa^{-1} \sim N^{\frac{1}{2}} \left(1- \frac{\lambda^2}{\lambda_c^2}\right) \text{ .}
\end{equation*}

\section{The prismatic model; intermediate field method}
\label{Def_mod}

In this section we 
recall from \cite{Giombi_2018}
the definition of the
%$O(N)^3$ invariant-tensor model 
prismatic tensor model
and  the 
implementation of the 
intermediate field method for this model.
%This section is inspired from the chapters 3 and 14 of the book \cite{tanasa_2021}.    

\begin{figure}[h!]
\centering
\includegraphics[scale=0.8]{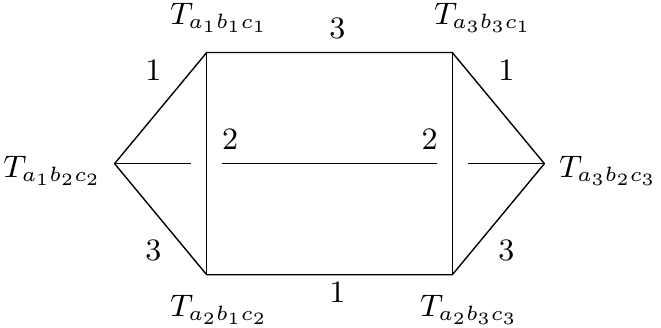}
\caption{The prismatic interaction}
\label{prismatic-fig}
\end{figure}

%During this internship, the action studied was the one having a single prismatic, sextic interaction:
Let a random tensor $T_{i_1 i_2 i_3}$ with three indices ranging from 1 to $N \in \mathbb{N}^*$. The group $O(N)^3$ acts 
%on it 
in the following way on the tensor field
\begin{equation}
T_{i_1 i_2 i_3} = O^{(1)}_{i_1 j_1} O^{(2)}_{i_1 j_1} O^{(3)}_{i_1 j_1} T_{j_1 j_2 j_3} \text{,}
\end{equation}
where the $O^{(k)}$ are independent orthogonal matrices.
The action of the model writes:
\begin{equation}
\label{actiune}
\begin{split}
S_N =- \frac{1}{2} & \sum_{i,j,k}  T_{ijk} T_{ijk}  \\
+ & \frac{t N^{-\alpha}}{6} \sum_{a_1, a_2, a_3, b_1,b_2, b_3, c_1, c_2,c_3 } T_{a_1 b_1 c_1} T_{a_1 b_2 c_2} T_{ a_2 b_1 c_2} T_{a_3 b_3 c_1} T_{a_3 b_2 c_3} T_{a_2 b_3 c_3} \text{,} \\ 
S_N =- \frac{1}{2} &\sum_{i,j,k} T_{ijk} T_{ijk} + \frac{t N^{-\alpha}}{6} I_p(T) \text{.} 
\end{split}
\end{equation}
Where 
\begin{equation}
I_p(T) = \sum_{a_1, a_2, a_3, b_1,b_2, b_3, c_1, c_2,c_3 } T_{a_1 b_1 c_1} T_{a_1 b_2 c_2} T_{ a_2 b_1 c_2} T_{a_3 b_3 c_1} T_{a_3 b_2 c_3} T_{a_2 b_3 c_3}
\end{equation}
is the \textbf{prismatic tensor invariant} whose graphical representation is given in Fig \ref{prismatic-fig}. Let us 
%emphasize 
note here that the prismatic interaction above is positively defined.

The partition function 
%and the $2-$point function are 
is then given by
\begin{equation}
\begin{split}
Z(t,N) &= \int [dT] e^{- \frac{1}{2} \sum_{i,j,k} T_{ijk} T_{ijk}+ \frac{t N^{-\alpha}}{6} I_p(T)}, \\
%G_2(t,N) &= \int [dT] T_{abc} T_{efg} e^{- \frac{1}{2} \sum_{i,j,k} T_{ijk} T_{ijk}+ \frac{t N^{-\alpha}}{6}}
%%\sum_{\{a_i, b_i, c_i \}} T_{a_1 b_1 c_1} T_{a_1 %b_2 c_2} T_{ a_2 b_1 c_2} T_{a_3 b_3 c_1} T_{a_3 %b_2 c_3} T_{a_2 b_3 c_3}}
%\text{.}
\end{split}
\end{equation}
where $[dT]$ is the tensor measure 
%(one integrates over the elements of the tensor $T$).
\begin{equation}
    [dT]=%\frac{1}{(2 \pi)^{N^3/2}}
    \prod_{i,j,k=1}^N d T_{ijk}.
\end{equation}

\begin{figure}[h!]
\centering
\includegraphics[scale=0.8]{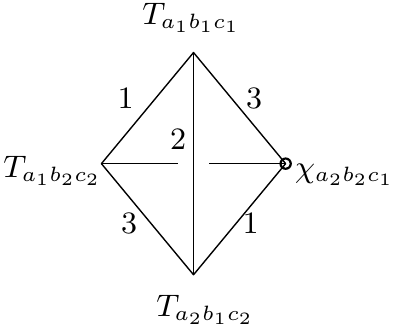}
\caption{The tetrahedric interaction}
\label{tetrahedric-fig}
\end{figure}

Using the intermediate field method, the
sextic interaction above can be reduced to a 
quartic interaction \cite{Lionni_2016}. 
Recall that
\begin{equation}
\label{intermediate}
\begin{split}
e^{ \frac{t N^{-\alpha}}{6} I_p(T)} = \int &\frac{[d \chi]}{(2 \pi)^{N^3/2}} e^{-\frac{1}{2} \sum_{i,j,k=1}^N \chi_{ijk} \chi_{ijk}} \\
&e^{\sqrt{\frac{2 t N^{-\alpha}}{6}} \sum_{a_1,a_2,b_1, b_2,c_1, c_2=1 }^N T_{a_1 b_1 c_1} T_{a_1 b_2 c_2} T_{a_2 b_1 c_2} \chi_{a_2 b_2 c_1}} \text{,} \\
= \int &\frac{[d \chi]}{(2 \pi)^{N^3/2}} e^{-\frac{1}{2} \sum_{i,j,k=1}^N \chi_{ijk} \chi_{ijk} + \sqrt{\frac{2 t N^{-\alpha}}{6}} \tilde{I}_t(T,\chi)}  \text{,}
\end{split}
\end{equation}
where  
\begin{equation}
 \tilde{I}_t(T,\chi) =  \sum_{a_1,a_2,b_1, b_2,c_1, c_2=1 }^N T_{a_1 b_1 c_1} T_{a_1 b_2 c_2} T_{a_2 b_1 c_2} \chi_{a_2 b_2 c_1} \text{.}
\end{equation}
The interaction obtained by using this intermediate field method is therefore a tetrahedron-type interaction coupling three $T$ fields and one intermediate field $\chi$. The graphical representation of this interaction is given in Fig \ref{tetrahedric-fig}.  

The partition function can therefore be written as
%begin{equation}
%\begin{split}
%Z = \int [dT] [d \chi ] &e^{-\frac{1}{2} \sum_{i,j,k=1}^N (T_{ijk} T_{ijk}+ \chi_{ijk} \chi_{ijk})} \\
%&e^{ \sqrt{\frac{2 t N^{-\alpha}}{6}} \sum_{a_1,a_2,b_1, b_2,c_1, c_2=1 }^N T_{a_1 b_1 c_1} T_{a_1 b_2 c_2} T_{a_2 b_1 c_2} \chi_{a_2 b_2 c_1}} \text{,} \\
% = \int [dT] [d \chi ] &e^{-\frac{1}{2} \sum_{i,j,k=1}^N (T_{ijk} T_{ijk} + \chi_{ijk} \chi_{ijk})} \\
% &e^{\frac{ \lambda' N^{-\alpha'}}{4} \sum_{a_1,a_2,b_1, b_2,c_1, c_2=1}^N T_{a_1 b_1 c_1} T_{a_1 b_2 c_2} T_{a_2 b_1 c_2} \chi_{a_2 b_2 c_1}}
% \text{,}
%\end{split}
%\end{equation}

\begin{equation}
\begin{split}
Z(t,N) = \int [dT] \frac{[d \chi]}{(2 \pi)^{N^3/2}} &e^{-\frac{1}{2} \sum_{i,j,k=1}^N (T_{ijk} T_{ijk}+ \chi_{ijk} \chi_{ijk})+ \sqrt{\frac{2 t N^{-\alpha}}{6}} \tilde{I}_t(T,\chi)} \text{,} \\
 = \int [dT] \frac{[d \chi]}{(2 \pi)^{N^3/2}} &e^{-\frac{1}{2} \sum_{i,j,k=1}^N (T_{ijk} T_{ijk} + \chi_{ijk} \chi_{ijk})+ \frac{ \lambda' N^{-\alpha'}}{4} \tilde{I}_t(T,\chi)}
 \text{,}
\end{split}
\end{equation}
where we made a redefinition of the coupling constant $\frac{ \lambda' N^{-\alpha'}}{4} = \sqrt{\frac{2 t N^{-\alpha}}{6}}$. 
 The identification $2 \alpha' = \alpha$ indicates that the scaling of a prismatic interaction needs to be twice the one of a tetrahedron.
Recall that the single field  tetrahedron interaction is known to be equal to $\frac{3}{2}$
\cite{Carrozza_2016}.
%in the case of a single field interaction. 
One thus has, for the scaling parameter of the prismatic interaction:
\begin{equation}
    \alpha = 3.
\end{equation}

Expanding the interaction part of the action and using Wick's theorem, one can write the partition function as a formal power series in the coupling constant such that:
\begin{equation*}
    Z(t,N) = \sum_{n_p \in \mathbb{N}} \sum_{\mathcal{G} \in \mathbb{G}_{n_p}} A_{\mathcal{G}} t^{n_p}
\end{equation*}
Where the second sums runs over the set of vacuum $4-$regular properly$-$edge$-$ colored graphs such that the subgraph
obtained by removing all the feynman edges\footnote{ The "feynman" edges, obtained from Wick's theorem, are given the color 0 in the graphs.}  is a disjoint union of either $n_p$ prismatic vertices, if no intermediate field was used, or $2 n_p$ tetrahedric vertices, if one was introduced. The parameter $A_{\mathcal{G}}$ is the amplitudes of the graph $\mathcal{G}$.
In the sequel, when working with prismatic vertices, it will be stated that the analysis is performed in the {\bf prismatic representation}. When working with the intermediate field representation, hence with tetrahedron interactions, we will refer to it as the {\bf tetrahedric representation}.

%As the intermediate fields cannot be present as an external leg and can only be Wick contracted with another auxiliary field, the only graphs in the tetrahedric representation are the ones built of pairs of tetrahedra interaction connected by an intermediate field propagator. In fact, one has $\lambda'^2 \propto \lambda$, implying that the order $n$ coefficient in a $\lambda$ expansion is equivalent to an order $2n$ coefficient in a $\lambda'$ expansion.

\begin{figure}[h!]
\centering
\includegraphics[scale=0.8]{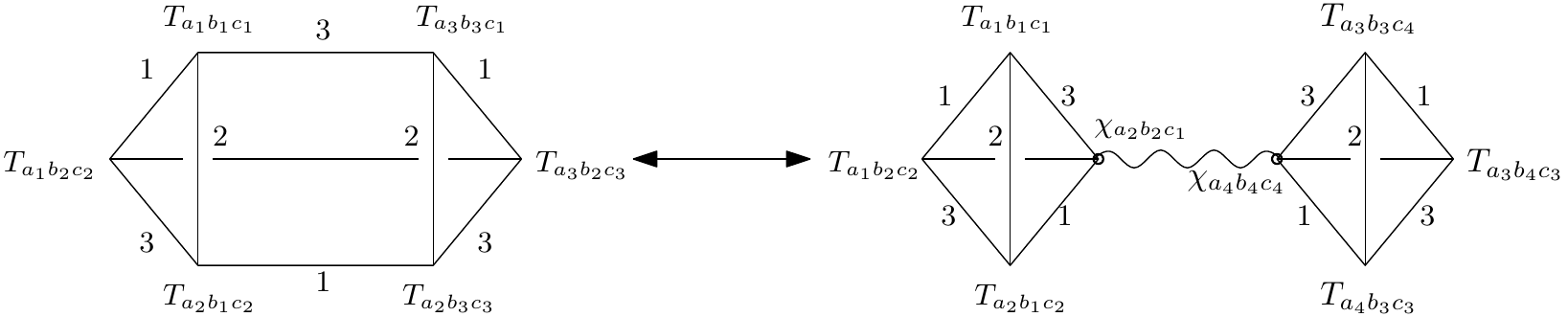}
\caption{The intermediate field method for the prismatic tensor interaction.}
\label{intermediate-fig}
\end{figure}

Using the intermediate field method described above, any prismatic vertex in a Feynman graph can be split into two tetrahedra vertices connected by an intermediate field propagator $\chi$ as in Figure \ref{intermediate-fig}, where the field $\chi$ is represented by a curly line.

\begin{figure}[h!]
\centering
\includegraphics[scale=0.80]{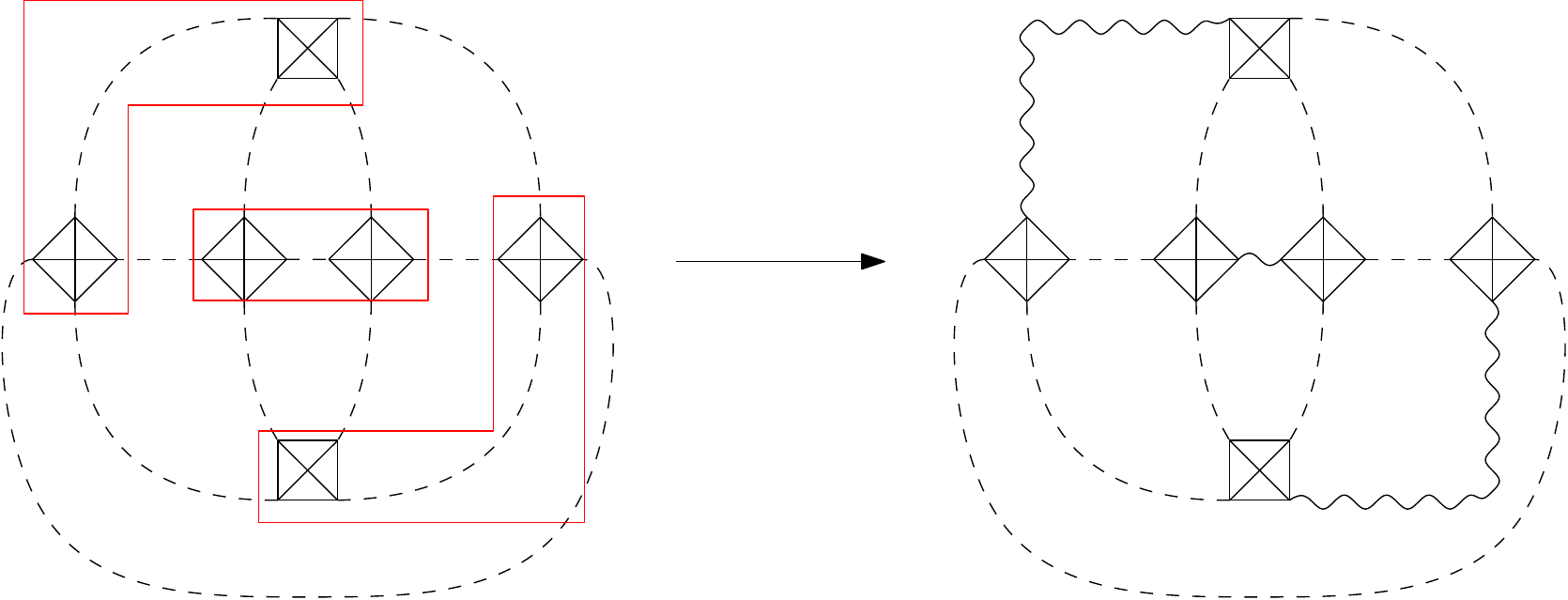}
\caption{On the left, a graph in the tetrahedric model and one of its perfect pairing. On the right is the corresponding graph in the tetrahedric representation.}% of the prismatic model.}
\label{dimere}
\end{figure}

\bigskip
Let us also notice that the tetrahedric representation can be linked to dimer models of the graphs present in the tetrahedric model.
The dimer model is the study of the perfect matchings of a graph. 
{Recall that a perfect matching of a graph is a configuration where all its vertices are paired with exactly one of their connected neighbours} \cite{https://doi.org/10.48550/arxiv.math/0310326}. As each vertex possesses a single $\chi$ node, connecting two vertices with a $\chi$ propagator  is equivalent to pairing them. Therefore, by considering the dimers of all the graphs in the tetrahedric model, one can derive all the graphs present in the tetrahedric representation of the prismatic model \footnote{In a more general setting, replacing one of the fields in a single field interaction by an intermediate field is equivalent to consider the dimers of the graphs present in the original model.}. An example of the perfect pairing of a graph in the tetrahedric model and its equivalent in the tetrahedric representation is given in Figure \ref{dimere}.

\bigskip

One can prove that the Feynman amplitude of a vacuum graph is proportional to
\begin{equation}
A_{\mathcal{G}} \propto 
%N^{3 - \sum_{l=1,2,3} \frac{k_l}{2}} 
N^{3 - \omega (\mathcal{G})}  \text{,}
\end{equation}
where $\omega(\mathcal{G})$ is a parameter that we call the degree.

From each graph $\mathcal{G}$, one can define three jackets of color $l$ by erasing the edges of the corresponding color in $\mathcal{G}$. %Applying an untwisting procedure on the vertices, 
These jackets can be treated as ribbon graphs with non orientable genus $k_l$ (see \cite{Dartois_2013}). Using Euler characteristic formula, one can find an expression for $\omega(\mathcal{G})$: 
\begin{equation}\begin{split}
\omega (\mathcal{G}) = \sum_{l=1,2,3} \frac{k_l}{2} &= 3 + 3 n_p - f_{\mathcal{G}} = 3 + \frac{3}{2} n_t -f_{\mathcal{G}} \text{ .}
\end{split}
\label{degree_eq}
\end{equation}
 In the equation above, $f_{\mathcal{G}}$ is the number of faces of the graph $\mathcal{G}$.

\medskip

From the partition function one can define the free energy as 
\begin{equation*}
    F(t,N)= -\frac{1}{N^3} \text{ln}(Z) \text{ .}
\end{equation*}
and the $2-$point function as
 \begin{equation*}
     \langle T_{abc} T_{a'b'c'} \rangle = G_{2} (t) \delta_{a a'} \delta_{b b'} \delta{c c'} \text{ ,}
 \end{equation*}
where $G_{2}(t) =\frac{1}{N^3} \langle \sum_{abc} T_{abc} T_{abc} \rangle$. 
Similarly to the partition function, $G_{2}(t)$ admits an expansion in terms of vacuum rooted graphs. Vacuum rooted graphs are vacuum graphs with a marked edge called the root. 
{When studying the $2-$point function in section \ref{sec_DS}, it will be understood that we work with rooted vacuum graphs.}

To obtain the free energy $F(t,N)$ from the $2-$point function, one can introduce a new parameter $s$ such that $t=\tilde t s^3$ and a rescaled field $\tilde T$ such that $T_{abc} = \frac{1}{\sqrt{s}} \tilde T_{abc}$. This introduces a parameter 
multiplying 
the quadratic part of the prismatic action. Denoting by a tilde the quantities with $s$ introduced, the action rewrites 
\begin{equation*}
    \tilde S_N=- \frac{1}{2} s^{-1} \sum_{i,j,k} \tilde T_{ijk} \tilde T_{ijk} + \frac{\tilde t N^{-3}}{6} I_p(\tilde T)  
\end{equation*}
and the free energy writes 
\begin{equation*}
    \tilde F (s, \tilde t,N) = -\frac{1}{N^3} \text{ln} \left(\int [d \tilde T] e^{- \frac{1}{2} s^{-1} \sum_{i,j,k} \tilde T_{ijk} \tilde T_{ijk} + \frac{\tilde t N^{-3}}{6} I_p(\tilde T)} \right) \text{ .}
\end{equation*}
The $2-$point function of the rescaled field is also given by
\begin{equation*}
    \begin{split}
    \tilde G_2 (\tilde t,s) = \sum_{a,b,c} \frac{1}{ \tilde Z(s,t,N)} \int [d \tilde T]  \frac{\tilde T_{abc} \tilde T_{abc}}{N^3} e^{- \frac{1}{2} s^{-1} \sum_{i,j,k} \tilde T_{ijk} \tilde T_{ijk} + \frac{\tilde t N^{-3}}{6} I_p(\tilde T)} \text{ ,}
    \end{split}
\end{equation*}
\begin{equation*}
    \tilde G_2 (\tilde t,s) = s G_2(\tilde t s^3) =   \frac{2 s^2}{ N^3} \frac{1}{\tilde Z(s,t,N)} \frac{\partial \tilde Z(s,t,N)}{\partial s} = -2 s^2 \frac{\partial \tilde F(s,t,N)}{\partial s} \text{ .}
\end{equation*}
Integrating $- \frac{G_2(\tilde t s^3)}{2s}$, from $s=0$ to $s=1$ at fixed $\tilde t$ gives then the original free energy since $\tilde F(s=1,\tilde t,N) = F(t,N)$ and $F(s=0,\tilde t,N) = 0 $:
\begin{equation*}
    F(t,N)  = - \int_{0}^{1} \frac{ds}{2s} G_2(\tilde t s^3),
\end{equation*}
where we have used the fact that $\tilde t = t$  when $s=1$.

\section{Leading order graphs in the large N limit}
\label{large N}

In this section, we study the $\frac{1}{N}$  expansion of the prismatic model and we explicitly exhibit the dominant graphs, in both the tetrahedric and prismatic representation.
Our study, as already mentioned in the Introduction, uses the intermediate field method.

\subsection{Leading order graphs in the tetrahedric representation}
\label{secLo}

%, first defined for colored tensor models \cite{Gurau_2012}. The degree $\omega$ is the parameter controlling the $\frac{1}{N}$ expansion of random tensor models. This parameter is not a topological invariant, this marks an important difference with random matrix models where the expansion is controlled by the genus of the graphs $k_{\mathcal{G}}$ \citep{DiFrancesco:2004qj}.

\begin{figure}[h!]
\centering
\includegraphics[scale=1]{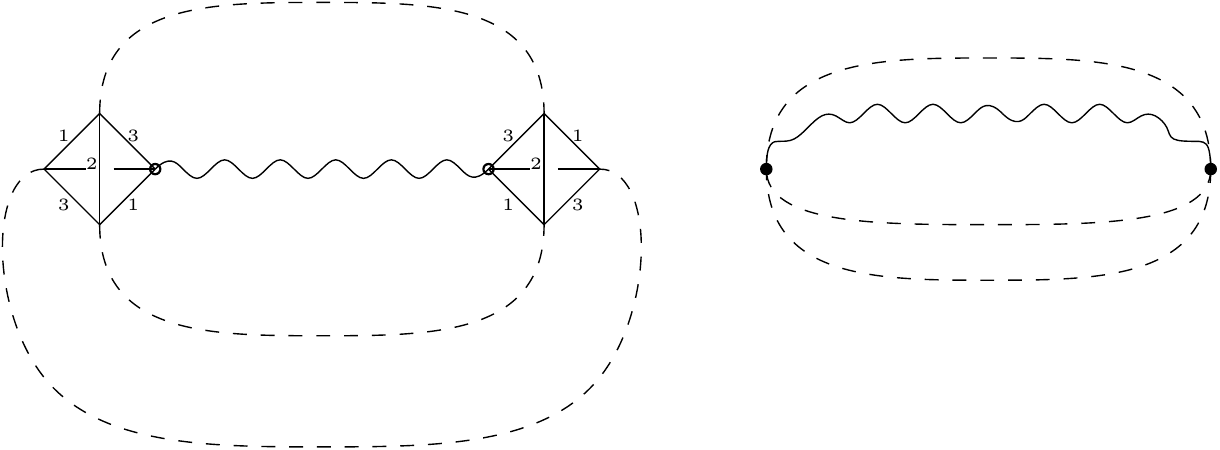}
\caption{On the left, the vacuum elementary melon in the tetrahedric representation. On the right, a simplified graphical representation of the graph where the tetrahedric vertices are replaced by black dots.}
\label{elementary}
\end{figure}

\begin{figure}[h!]
\centering
\includegraphics[scale=1.2]{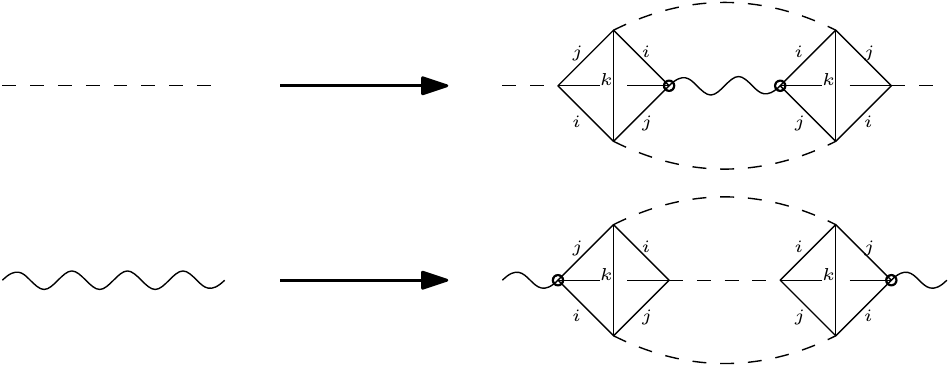}
\caption{The melonic insertions on the two types of propagators.}
\label{melonic_insertion}
\end{figure}

Equation \eqref{degree_eq} above implies that the degree of the prismatic model is a non-negative integer.
Hence, the LO Feynman graphs of the model are thus the ones satisfying the condition:
\begin{equation}
\omega (\mathcal{G}) = 0 \text{.}
\end{equation} 
Let us first identify these graphs in the tetrahedric representation. 
As usual in the tensor model literature, we call these graphs melonic graphs.
 One can prove that they are built by starting from the vacuum elementary melon shown in Figure \ref{elementary} and by recursively inserting one of the elementary 2-points melons on arbitrary edges. The insertions possible on the two types of edges are shown in Figure  \ref{melonic_insertion}.

%The proof that the melons are the only degree 0 graphs in the quartic model is well known. Different variants of it were proposed for different tensor models as in \cite{Dartois_2013}, \cite{Carrozza_2016} or \cite{https://doi.org/10.48550/arxiv.1808.09434}. In the sequel, we adapt the proof of \cite{https://doi.org/10.48550/arxiv.1808.09434} . 

\begin{figure}[h!]
\centering
\includegraphics[scale=1.2]{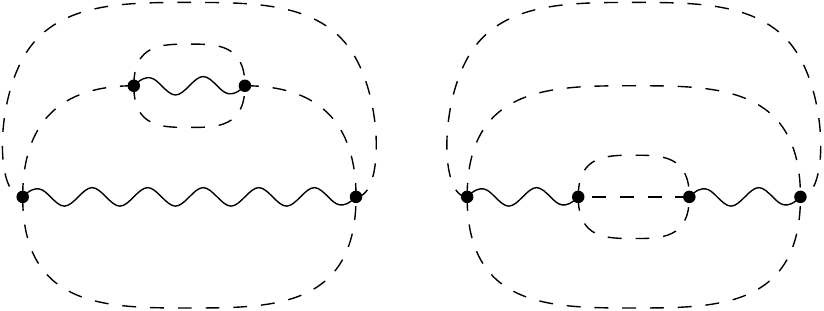}
\caption{Example of melonic graphs.}
\label{melonic_example}
\end{figure}

Let us give in Fig. \ref{melonic_example} an example of melonic graphs obtained {\it via} melonic insertions on a $T$ propagator, on the LHS, and on a $\chi$ propagator on the RHS.

\subsection{Leading order graphs in the prismatic representation}

\begin{figure}[h!]
\centering
\includegraphics[scale=1.0]{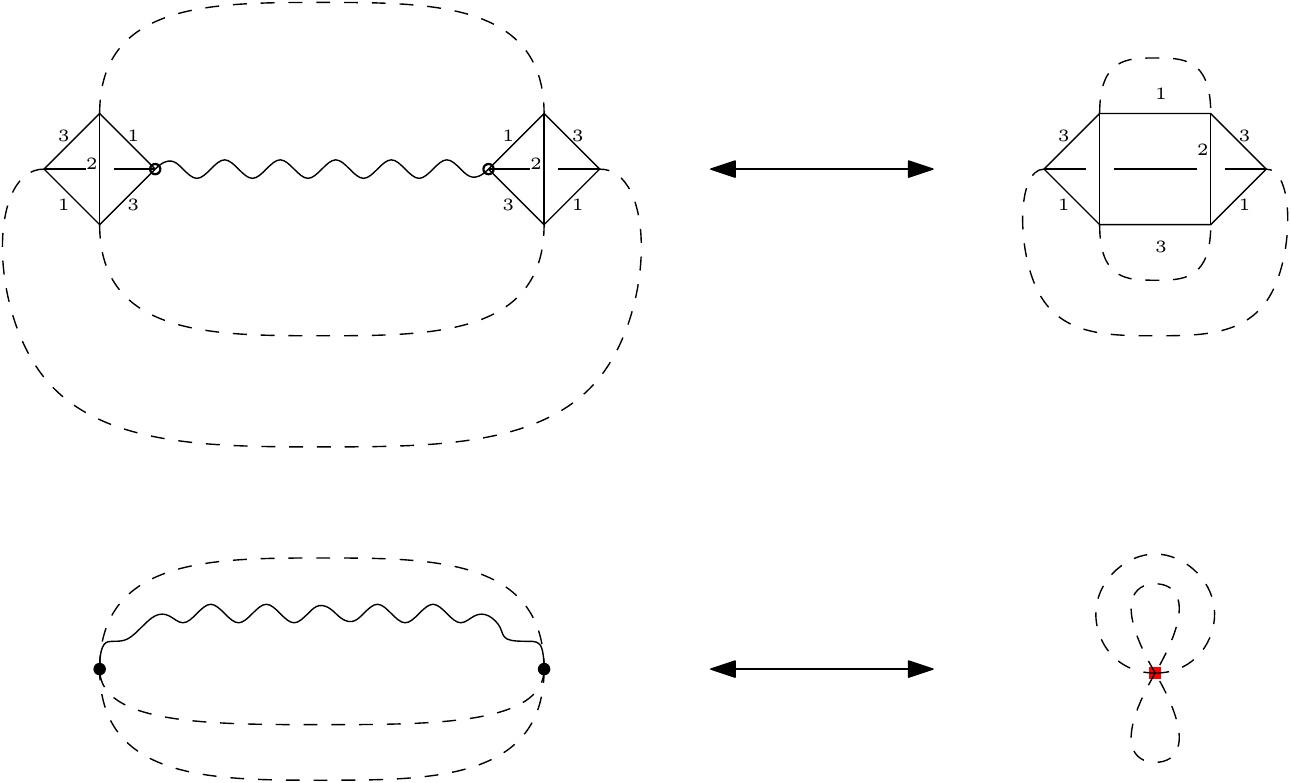}
\caption{On the left, the elementary vacuum melon in the two graphical representations. On the right, their prismatic equivalent. On the bottom right, the prismatic vertex is represented by a red square.}
\label{prism_lo}
\end{figure}

%The LO graphs can also be exhibited in the prismatic representation by replacing back every pair of tetrahedra connected by a $\chi$ propagator by a prism. 
In the prismatic representation, the elementary melon of the tetrahedric representation becomes a triple tadpole (see Figure \ref{prism_lo} for its various graphical representations). 
We refer to this triple tadpole as the {\it elementary triple tadpole}. 

\begin{figure}[h!]
\centering
\includegraphics[scale=0.9]{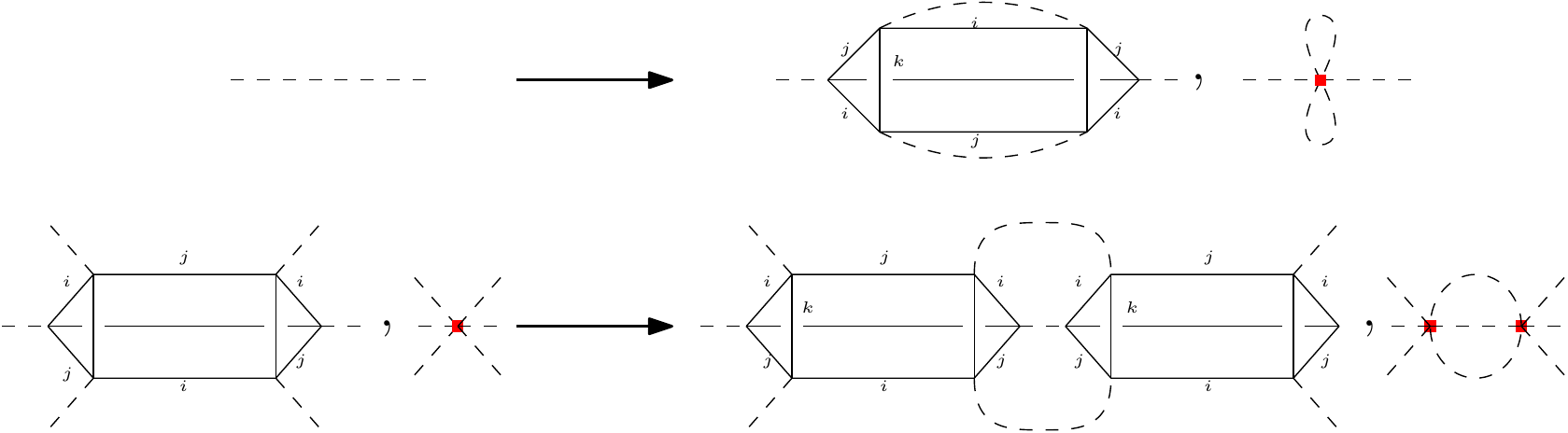}
\caption{The prismatic equivalent of the two melonic insertions.} 
%The top one is the equivalent of the melonic insertion on a $T$ propagator and the bottom one the equivalent of the one on a $\chi$ propagator.}
\label{prism_insertion}
\end{figure}

Let us now explain what the two melonic insertions in the tetrahedric representation become in this prismatic representation.
The insertion on a $T$ propagator is equivalent to the insertion of a $2-$point double tadpole (which is obtained by "cutting" an edge of the elementary triple tadpole).
The insertion on a $\chi$ propagator is equivalent to an insertion at the level of a prismatic vertex.
One needs to "split" a prismatic vertex into two prismatic vertices which are linked by $T$ propagators (see Figure \ref{prism_insertion}).

\begin{figure}[h!]
\centering
\includegraphics[scale=0.8]{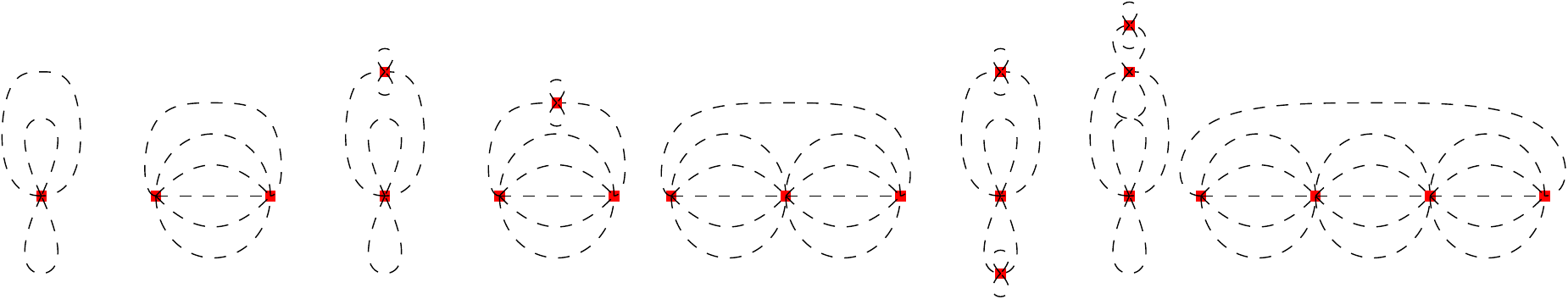}
\caption{Examples of the LO graph in the prismatic representation.} 
%The prisms are represented by red squares and the propagators by normal lines.}
\label{exemple}
\end{figure}

Let us give in Figure \ref{exemple} some examples of LO graphs in the prismatic representation. The first graph is the elementary triple tadpole. The second graph is obtained by a vertex splitting of the elementary triple tadpole vertex. 
%This second graph can be seen as an elementary melon graph 
The third graph is obtained by inserting a $2-$point double tadpole on one of the edges of the first graph (the elementary triple tadpole) and so on.

\section{Generating functions of leading order graphs}
\subsection{Generating functions of melonic graphs}

In this subsection we analyse the generating functions of melonic graphs in the tetrahedric representation, find its singularities and the behaviour of this generating function close to the singularity.

%Recall that vacuum melonic graphs were introduced in Section \ref{secLo}. 
%2-point melonic graphs can be defined similarly by inserting melons on the 2-point elementary melon instead of the vacuum one. 

\begin{figure}[h!]
\centering
\includegraphics[scale=0.85]{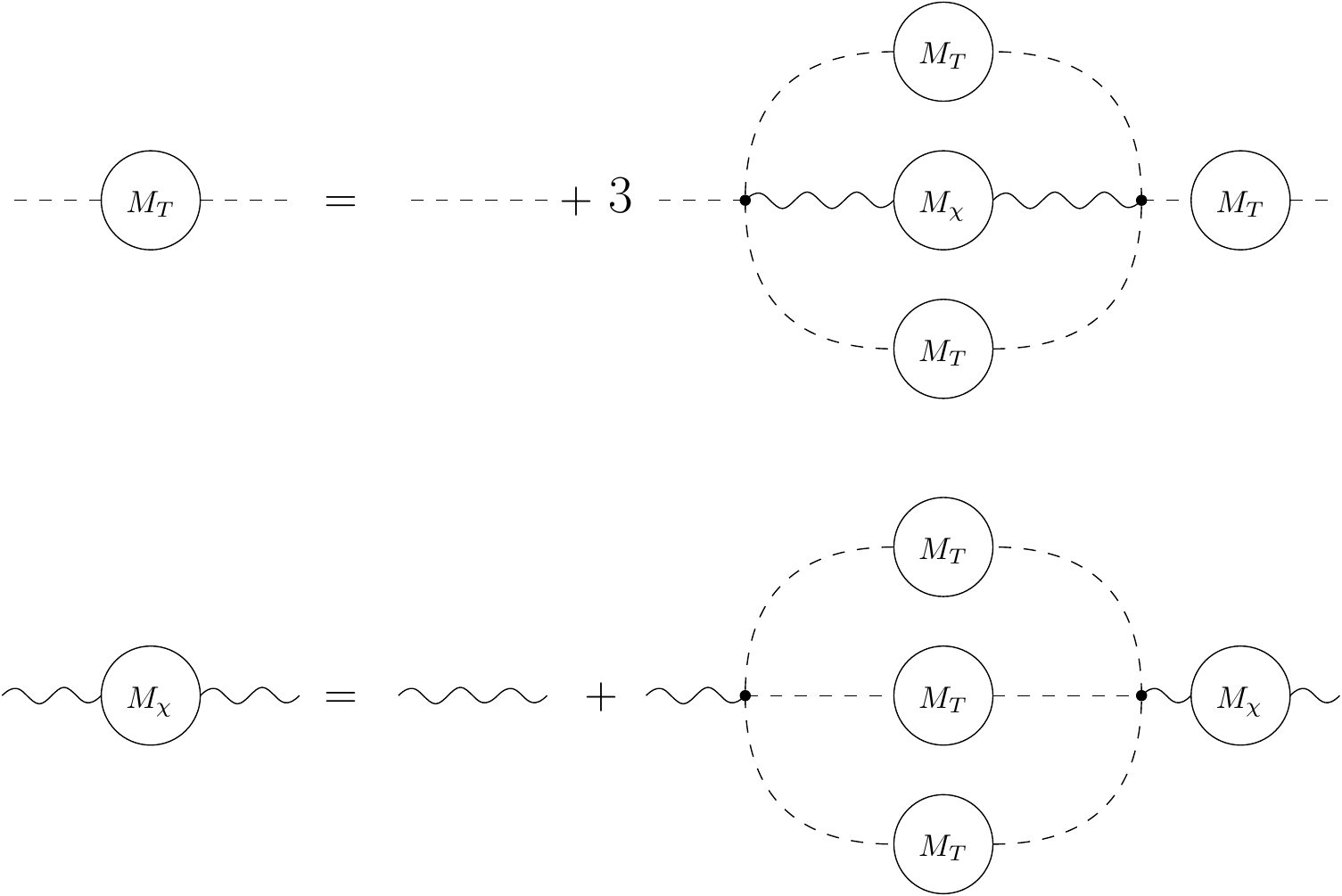}
\caption{LO $2-$point functions in the terahedric representation.}
%Structure of the melonic graphs in the case of a $T^3 \chi$ tetrahedric interaction.}
\label{generating_melon}
\end{figure}

As there are two types of $2-$point functions (associated with the two propagators), we denote by $M_T(t)$ and $M_\chi(t)$ the generating functions of the two sets of corresponding melon graphs. 
%As there are two types of $2-$point melon graphs, 
%there are two sets of graphs whose structures 
This is given in Figure \ref{generating_melon}.

One has
\begin{equation}
\begin{split}
M_T(t) &= 1+ 3 t M^3_T(t) M_\chi (t) \text{,} \\
M_\chi(t) &= 1 + t M^3_T (t) M_\chi (t) \text{,}
\end{split}
\end{equation}
where the factor $3$ in the first equation above counts the three different choices of placing a $\chi$ propagator connecting the two vertices of the melonic graph. Substituting $M_\chi (t)$ in the first equation gives

\begin{equation}
1+t( M^4_T (t) +2 M^3_T (t)) - M_T (t)= 0 \text{.}
\label{Melonic_SD}
\end{equation}
Finding the general solution of such equation is an involved task. 
%However, information about the solution can be derived without solving it. For example 
However, one can derive its singularities, \textit{i. e.} the points where the solution stops being analytic. 

%To understand how to find these points, o
One can consider the more general problem given by
\begin{equation}
F(M_T,t) = 0,
\end{equation}
where one needs to find $M_T(t)$. Using the implicit function theorem \cite{10.5555/1506267}, one can write 
\begin{equation}
\frac{d M_T (t)}{dt} = - \frac{\partial F/ \partial t}{\partial F / \partial M_T} \text{.}
\end{equation}
This implies that $M_T (t)$ ceases to be analytic when $\frac{\partial F}{\partial M_T} = 0$ as its derivative blows up. 

In the case of the melons, one has $F(M_T,t) = 1 +t ( M_T (t)^4 + 2 M_T (t)^3) - M_T (t)$.
%, which gives the system of equations:
One then has the following system of equations (we denote $M_T (t_c)$ by $M_{T,c}$):
\begin{equation}
\begin{split}
M_{T,c} &= 1 + t_c (M_{T,c}^4 + 2 M_{T,c}^3 ) \text{,} \\
1 &= t_c ( 4 M_{T,c}^3 + 6 M_{T,c}^2) \text{.}
\end{split}
\end{equation}
The second equation implies $t_c = \frac{1}{4 M_{T,c}^3+ 6 M_{T,c}^2}$, which can be substituted in the first one to give
\begin{equation}
M_{T,c} = 1 + \frac{M_{T,c}^4 + 2 M_{T,c}^3}{4 M_{T,c}^3 + 6 M_{T,c}^2} \text{.}
\end{equation}
After some algebra, one finds $M_{T,c}^2 = 2$. The singularities are then the points for which $M_T = \pm \sqrt{2}$ at $t_c = \frac{1}{12 \pm 8 \sqrt{2}}$. Recall that Pringsheim's theorem (see for example \cite{10.5555/1506267}) implies that the singularity closest to $t=0$ defines the radius of convergence of the series expansion of $M_T(t)$ around the origin. This point is the dominant singularity.
%, it is the most relevant to study when working with series expansion of generating functions. 
As $\frac{1}{12 + 8 \sqrt{2}} < \frac{1}{12 - 8 \sqrt{2}}$, one has 
\begin{equation}
\label{tc}
    t_c=\frac{1}{12 + 8 \sqrt{2}} \mbox{ and } M_T (t_c)= \sqrt{2},
\end{equation}
at the dominant singularity. 

Let us now find the behavior of the function $M_T (t)$ near this dominant singularity. 
In order to do this, one needs to
Taylor expand the function $F(M_T,t)$ near a critical point.

Assuming that $ \frac{\partial^2 F}{\partial M^2_T} \neq 0$, 

and using the conditions $F(M_T,t) =0$, $\frac{\partial F}{\partial M_T}\vert_{t_c} =0$
and  
since $\frac{\partial F}{\partial t^2} =0$, one has:
\begin{equation}
\begin{split}
\frac{\partial F}{\partial t}|_{t_c} (t_c-t) + \frac{\partial^2 F}{\partial t \partial M_T}|_{t_c} (t_c-t) &(M_T(t) - M_{T,c}) + O(|t_c-t|^2) \\
& = \frac{\partial^2 F}{\partial M_T^2}|_{t_c} \frac{1}{2} (M_T(t) - M_{T,c})^2 \text{.}
\end{split}
\end{equation}
In the limit $t \rightarrow t_c$ (keeping only the LO term in the expansion), one has:
\begin{equation}
\label{sol}
M_T (t) \stackrel{t \rightarrow t_c}{\sim} M_{T,c} \pm \sqrt{2 \frac{\partial F / \partial t |_{t_c}}{\partial^2 F / \partial M_T^2 |_{t_c}}} \sqrt{t_c-t} \text{.}
\end{equation}

Since $M_T (t)$ is the generating function of melonic graphs, this implies that $M_T(t)$ is an increasing function in $t$ and this further implies that one needs to choose, amongst the two solutions \eqref{sol}, the solution with a negative sign.

Thus, the function $M_T (t)$ behaves, near the dominant singularity, as 
\begin{equation}
\begin{split}
&M_T(t) \stackrel{t \rightarrow t_c}{\sim} M_{T,c} - \sqrt{ \frac{M^4_{T,c}+ 2 M^3_{T,c}}{6 (M_{T,c}^2 + M_{T,c}) }} \sqrt{1-t/t_c} \text{,} \\
& M_T(t) \stackrel{t \rightarrow t_c}{\sim} M_{T,c} - K \sqrt{1-t/t_c} \text{.}
\end{split}
\end{equation}

Inserting the explicit expressions 
\eqref{tc}
of $t_c$ and $M_{T,c}$, this behavior writes as

\begin{equation}
M_T (t) \stackrel{t \rightarrow t_c}{\sim} \sqrt{2} - \sqrt{\frac{\sqrt{2}}{3} (1 - \frac{t}{12 + 8 \sqrt{2}})} \text{.}
\end{equation}

%\subsection{Enumeration of leading order graphs in the prismatic representation}

\subsection{Generating function of LO graphs in the prismatic representation}

%Having identified the LO graphs, we now enumerate them at each order in perturbation theory. 

In this subsection, we analyze the generating function of LO graphs in the prismatic representation and we give an enumerative combinatorics result for the number of such LO graph at an arbitrary order in perturbation theory.

Let us denote by $P$ the generating function of the $2-$point LO graphs. This expands as $P =  \sum_n a_n t^n$ 
%can be performed. 
%Because the coefficients/scaling of the interaction are chosen such that all LO graphs have the same amplitude, t
The coefficients $a_n$ thus give us the number of LO graphs at a given order $n$ in perturbation theory.
%enumerate the ones with $n$ vertices. 
In order to compute these coefficients, a closed equation for $P$ needs to be found. This equation can be obtained in a  diagrammatic way. 
%Since we found two spe
%given by a 2-point tadpole and a separation of a vertex, see Figure \ref{SD_LO}. 

\begin{figure}[h!]
\centering
\includegraphics[scale=0.9]{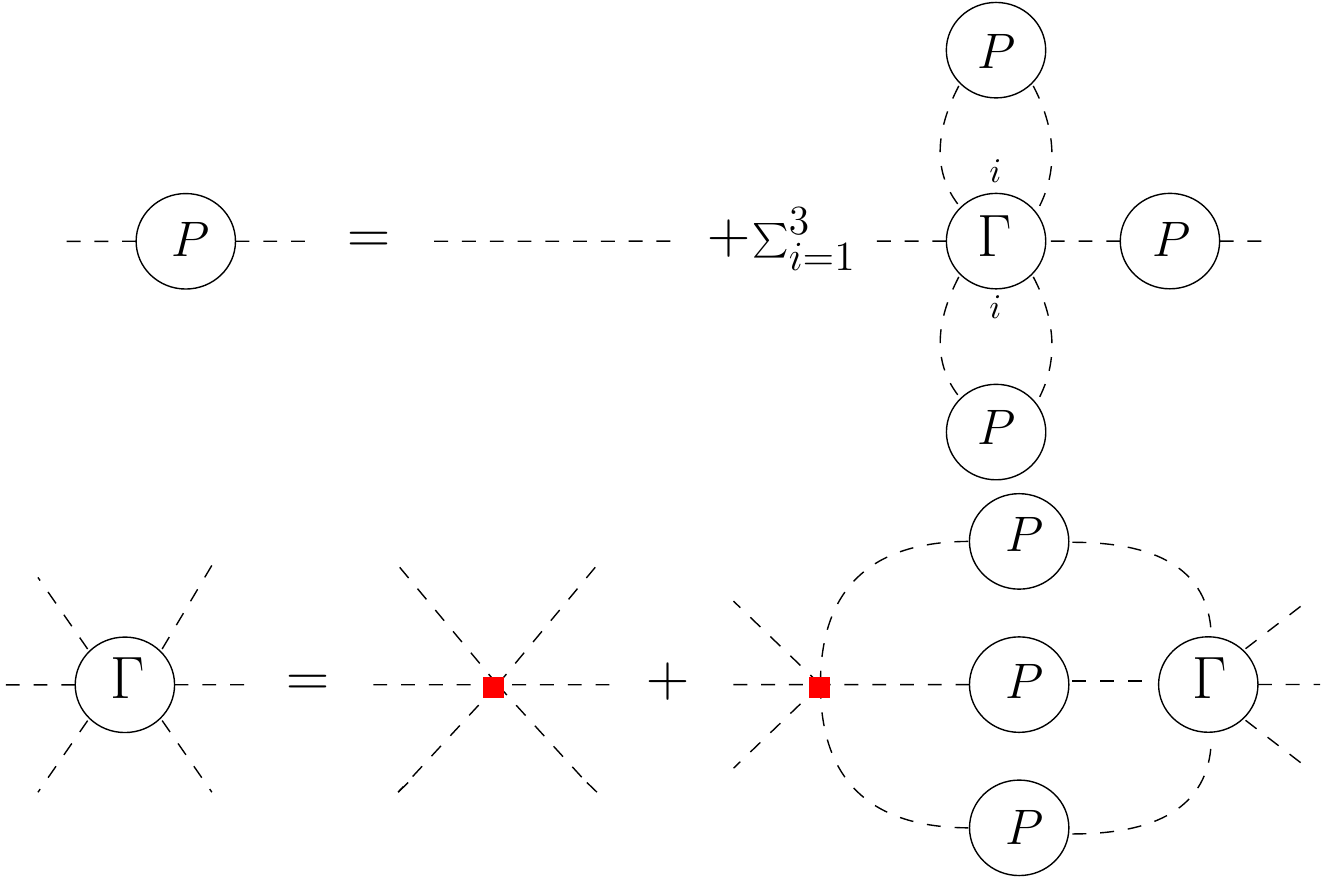}
\caption{Graphical representation of the equations of the generating function $P$.}
\label{SD_LO}
\end{figure}

Recall that we found two distinct insertions (one at the level of the propagator and one at the level of the prismatic vertex) which generate all the LO graphs in the prismatic representation. This translates in the diagrammatic equation represented in Fig. \ref{SD_LO}, for the $2-$point and $6-$point function. 

One has
\begin{equation}
\begin{split}
P &= 1+ 3 P^3 \Gamma \text{,}\\
\Gamma & = t + t P^3 \Gamma \text{.}
\end{split}
\end{equation}
%Eliminating $\Gamma$ gives 
%$P = 1 + \frac{3 P^3 t}{1- t P^3}$ which leads to
This leads to
\begin{equation}
P-1 = t ( P^4 + 2 P^3) \text{.}
\end{equation}
Changing variable to $y= P-1$ gives
\begin{equation}
y = t \big{(} (y+1)^4 + 2 (y+1)^3 \big{)} = t f(y) \text{.}
\end{equation}

Using Lagrange inversion theorem (see again the book \cite{10.5555/1506267}), the coefficient $a_n$ can be computed in the following way. Suppose that  $f(x)$ expands as $f(x) = \sum_{k \geq 0} f_k x^k$ and $f_0 \neq 0$. Then the equation $y= t f(y)$ admits a unique solution of the form $y(t)= \sum_{n \geq 1} y_n t^n$ with $y_n= \frac{1}{n} [x^{n-1}]f(x)^n$, where the notation $[x^m] g(x)$ denotes the order $m$ coefficient of $g(x)$. 

Recall that, in the case considered here, $f(y) = ( y+1)^4 + 2 (y+1)^3 = (y+1)^3(y+3)$.
This  implies that $y_n = \frac{1}{n} [x^{n-1}]\big{(} (x+1)^{3n} (x+3)^n \big{)}$. 
Using the binomial expansion, one can identify the coefficients as %one looks for as
%gives 
%\begin{equation}
%(x+1)^{3n} (x+3)^n = \sum_{ k \geq 0}^{3n} \sum_{j \geq 0}^n C_{3n}^k C_n^j x^{j+k} 3^{n-j}
%\end{equation}
%and replacing $j$ by $n-k-1$, $y_n$ reads
\begin{equation}
y_n = 
%\frac{1}{n} \sum_{k \geq 0}^{n-1} C_{3n}^k C_n^{n-k-1} 3^{k+1} = 
\frac{1}{n} \sum_{k \geq 0}^{n-1} \frac{3n!}{k! (3n-k)!} \frac{n!}{(n-k-1)!(k+1)!}3^{k+1} \text{.} 
\end{equation}
Since $y=P-1$, one thus has
\begin{equation}
P=1+ \sum_{n \geq 1}  \sum_{k \geq 0}^{n-1} \frac{3n!}{k! (3n-k)!} \frac{n!}{(n-k-1)!(k+1)!}3^{k+1} t^n \text{,}
\end{equation}
which finally leads to the identification of the number of LO graphs in the prismatic representation to be: $a_0=1$ and $a_n=y_n$, for any $n\in{\mathbb N}$.

\section{Diagrammatic Analysis}
\label{Diagrammatic Analysis}

In this section we introduce the diagrammatic tools needed to implement the double scaling limit of the model.

\subsection{Dipoles and their generating functions}
\label{Dipole_sec}

The notion of dipole was already introduced for 
the colored tensor model in \cite{Gurau_2011_2}, then for 
the MO tensor model in \cite{https://doi.org/10.48550/arxiv.1408.5725} and also for the 
$O(N)^3-$invariant quartic 
model in \cite{Bonzom_2022}. 
%It is possible to generalize them to a $T^3 \chi$ interaction by giving a different definition than the two previous papers.
In this subsection, we define a diagrammatic notion of dipole subgraph for the 
%$T^3 \chi$ model studied here.
model studied here, in the tetrahedric representation.

\medskip

A {\bf dipole} of color $i$ is a subgraph 
%(with possible non trivial melonic insertions on the edges) 
formed by two tetrahedric vertices connected by two parallel edges such that the subgraph has a face of length two and of color $i$. 

\medskip

%can be defined as any 4-point graph that becomes a face of length 2 and color $i$ incident to two distinct vertices, once all their melonics two points subgraph have been removed. 

\begin{figure}[h!]
\centering
\includegraphics[scale=1.1]{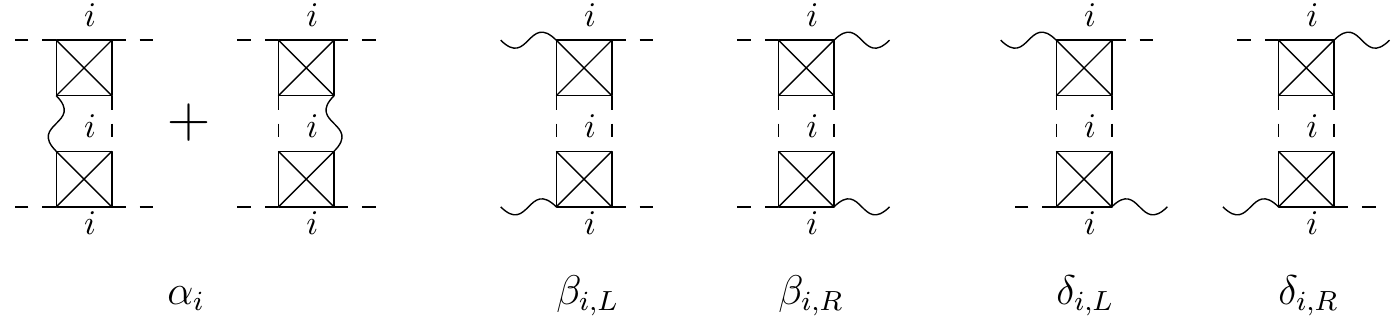}
\caption{The five types of dipoles.}
\label{dipoles}
\end{figure}

%With a $T^3 \chi$ tetrahedric interaction 
For each given color $i$, there are five different types of dipoles. We call these types of dipoles types $\alpha$, $\beta_L$, $\beta_R$, $\delta_L$ and $\delta_R$, see Fig. 
\ref{dipoles}.

\begin{figure}[h!]
\centering
\includegraphics[scale=0.8]{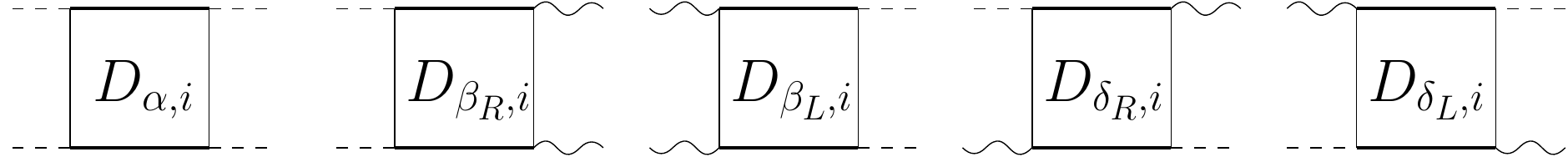}
\caption{The five types of dipole-vertices.}
\label{Dipoles}
\end{figure}

To simplify the graphical representation, the dipoles can be replaced by the dipole-vertices shown in Figure \ref{Dipoles}.

Let us now exhibit the generating functions for the dipoles.
One needs to insert the generating functions of melons on each edge and its square root on its half edges. 
%One needs to consider the insertion of melons on each edge of a dipole and needs to count the melons on the external edge with a power $\frac{1}{2}$.  
This way of counting the melonic insertions on the edges allows us to avoid counting twice the melons when there are two adjacent dipoles.

Thus, the generating functions of the five types of dipoles write:
\begin{equation}
\begin{split}
& D_{\alpha} = 2 t M_T(t)^3 M_\chi = \frac{2}{3} (M_T(t)-1)  \text{,}\\
& D_{\beta_L} = D_{\beta_R}= D_{\beta} = t M_T(t)^3 M_\chi(t) = \frac{1}{3} (M_T(t) -1)  \\
&D_{\delta_L}= D_{\delta_R}= D_\delta = t M_T(t)^3 M_\chi(t) = \frac{1}{3} (M_T(t) -1) \text{.}
\end{split}
\end{equation}

%Note that the color of a dipole doesn't play a role in its generating function. 

Note that the generating function of a dipole 
does not depend on the color. Notice also that the generating functions of the dipoles of type $\beta_L$ and $\beta_R$ as well as the generating functions of the dipoles of type $\delta_L$ and $\delta_R$ are the same. In the following, we will differentiate the generating function of these dipoles when writing the structure of the chains (in Section \ref{chains_sec}). However, when writing the contribution of the dominant graphs to the $2-$point function (in Section \ref{sec_DS}), we will simply denote them by $D_\beta$ and $D_\delta$. This allows us to clarify the structure of the chains while keeping a compact expression for the contribution of the dominant graphs.  

From the expression of the different generating functions above, one can find that the critical points of the dipole
generating functions
are identical to the ones of the generating functions of the melon graphs.

\subsection{Chains and their generating functions}
\label{chains_sec}

{\bf Chains} are defined in an analogous way as in 
 \cite{Bonzom_2022} \cite{https://doi.org/10.48550/arxiv.1408.5725}
to be the $4-$point graphs obtained by connecting at least two dipoles and by matching one side of a dipole to the corresponding side of the next dipole in the chain.

A chain is said to be of length $\ell$ if it contains $\ell$ dipoles. Each chain of length $\ell$ contains subchains of length $2 \leq \ell' \leq \ell$. A chain is therefore said to be maximal if it cannot be included in a longer chain of the graph. 

Note that changing the length of a chain doesn't change the degree of a graph. 
This can be proven by induction on the length of the chain, by inserting a dipole in a chain of length $k$, and by counting carefully the number of faces and vertices.
%and applying the induction hypothesis.  

Using the previous five types of dipoles, there are eight different possibilities for the external edges of a chain. 
We thus have eight types of chains%, three of them having the labels $L$ or $R$. 
If all the dipoles of a given chain are of color $i$, the respective chain is said to be an {\it unbroken} chain of color $i$. 
If the dipoles of the chain do not have the same color, we say that the respective chain is  
{\it  broken}. 

\begin{figure}[h!]
\centering
\includegraphics[scale=0.9]{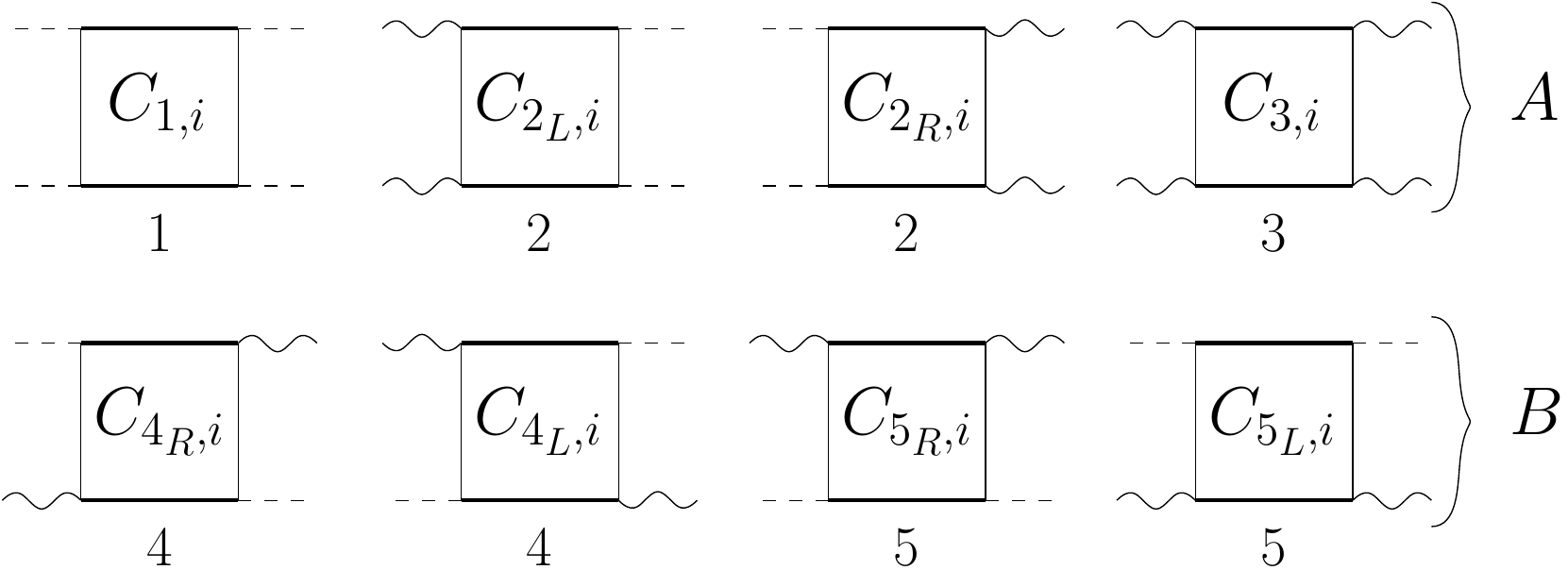}
\caption{All the different types of broken chain-vertices.}
\label{chains}
\end{figure}

Similar to the case of dipoles, in order to simplify the graphics, we represent unbroken chains of a given type 
%$s$ ($s=1,\ldots, 5$) 
and color $i$ by a chain-vertex $C$, while broken chains are represented by chain-vertices $B$. The different types of unbroken chain-vertices are shown in Figure \ref{chains}. 

Furthermore, we classify the chains into two families:
\begin{itemize}
\item Family $A$, if the external legs are the same on each side of the chain.
\item Family $B$, if the external legs are different on each side of the chain.
\end{itemize}  

%The two families are made of different arrangements of dipoles.

Let us now have a closer look at the dipole structure 
and the generating functions 
of each of these chains:
\paragraph{Family A:}
The building block of the chains in the family A are the dipoles of type $\alpha$, $\beta_L$ and $\beta_R$. If a dipole of type $\beta_R$ doesn't end a chain, it must be followed by a dipole of type $\beta_L$. 

\begin{figure}[h!]
\centering
\includegraphics[scale=1.0]{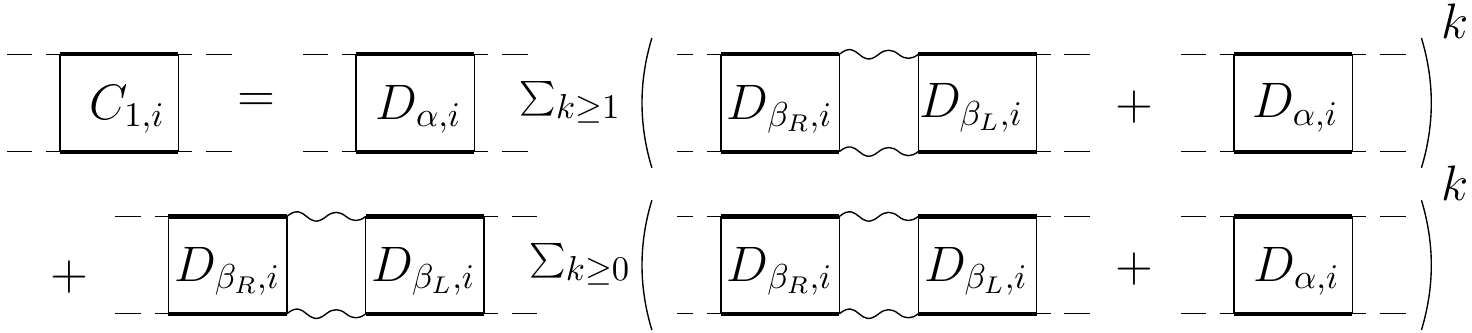}
\caption{The structure of the chains of type 1.}
\label{chain_type_1}
\end{figure}

\begin{figure}[h!]
\centering
\includegraphics[scale=0.7]{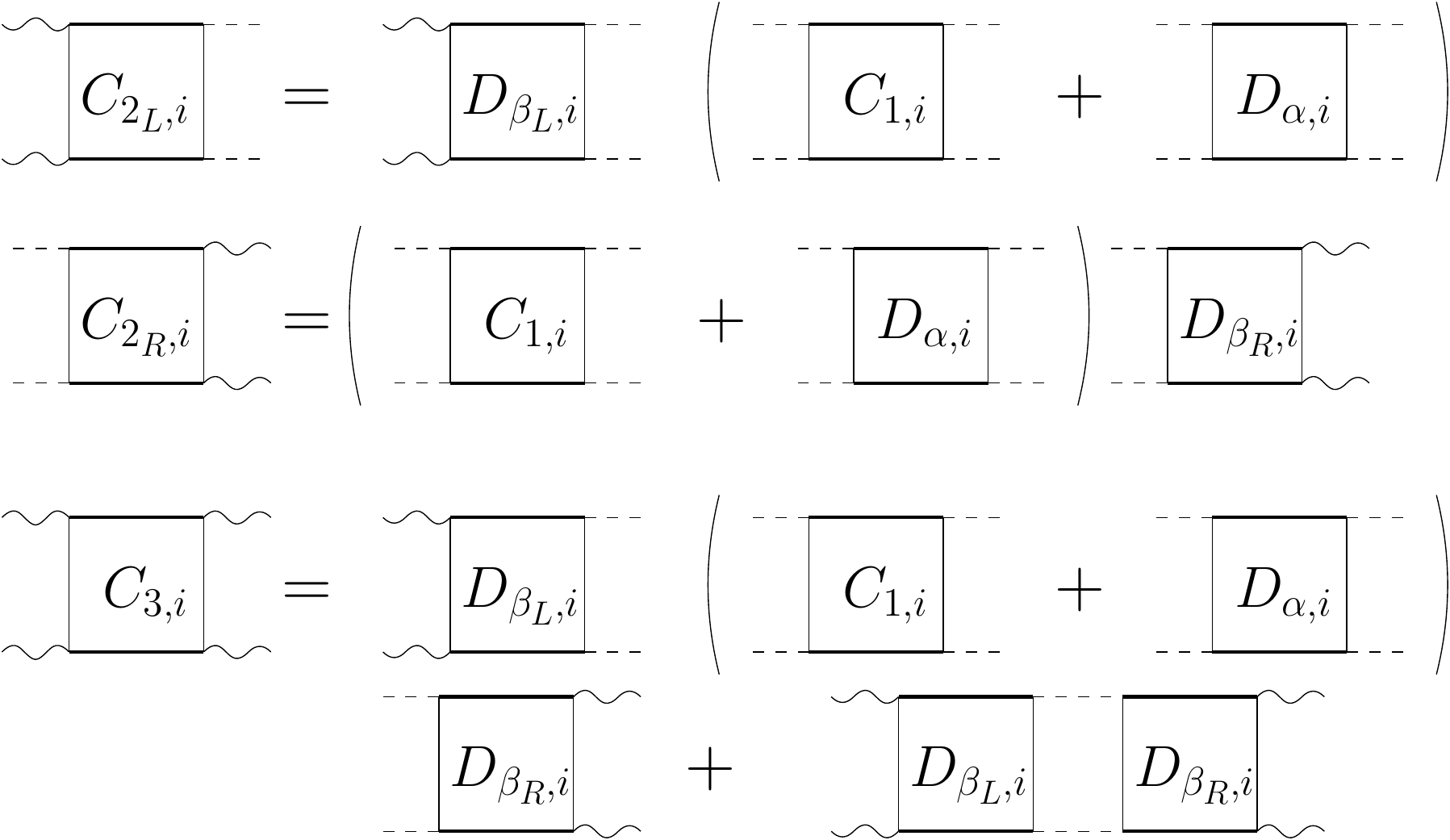}
\caption{The structure of the chains of type $2,3$.}
\label{family_A}
\end{figure}

%When looking closer to the chain of type $1$, a
All the possibilities of gluing up dipoles are given in Figure \ref{chain_type_1}. The other types of chains of the family $A$ follow directly from the type $1$ chain as shown in Figure \ref{family_A}. 

The generating functions are then derived as following:
\begin{equation}
\begin{split}
C_{1} &= D_{\alpha} \sum_{k \geq 1} (D_{\beta_R} D_{\beta_L} + D_\alpha)^k + D_{\beta_R} D_{\beta_L}  \sum_{k \geq 0} (D_{\beta_R} D_{\beta_L} + D_\alpha)^k  \text{,} \\
&= (D_{\alpha} + D_{\beta_R} D_{\beta_L}) \sum_{k \geq 0} ( D_\beta^2 + D_\alpha)^k - D_\alpha \text{,} \\
&= \frac{D_\alpha + D_{\beta_R} D_{\beta_L}}{1- ( D_\alpha + D_{\beta_R} D_{\beta_L})} - D_\alpha \text{,}
\end{split}
\label{chain_t1}
\end{equation}
and 
\begin{equation}
\begin{split}
C_{2_L} &=  D_{\beta_L} ( C_{1} + D_\alpha ) = D_{\beta_L} \frac{D_\alpha +  D_{\beta_R} D_{\beta_L}}{1- ( D_\alpha +  D_{\beta_R} D_{\beta_L})} \text{,}\\
C_{2_R} &=  D_{\beta_R} ( C_{1} + D_\alpha ) = D_\beta \frac{D_\alpha + D_{\beta_R} D_{\beta_L}}{1- ( D_\alpha + D_{\beta_R} D_{\beta_L})} \text{,}\\
C_{3} &= D_{\beta_L} (1+ C_{1} + D_\alpha ) D_{\beta_R} = \frac{D_{\beta_L} D_{\beta_R}}{1- (D_\alpha + D_{\beta_L} D_{\beta_R})} \text{.}
\end{split}
\end{equation}

\begin{figure}[h!]
\centering
\includegraphics[scale=0.8]{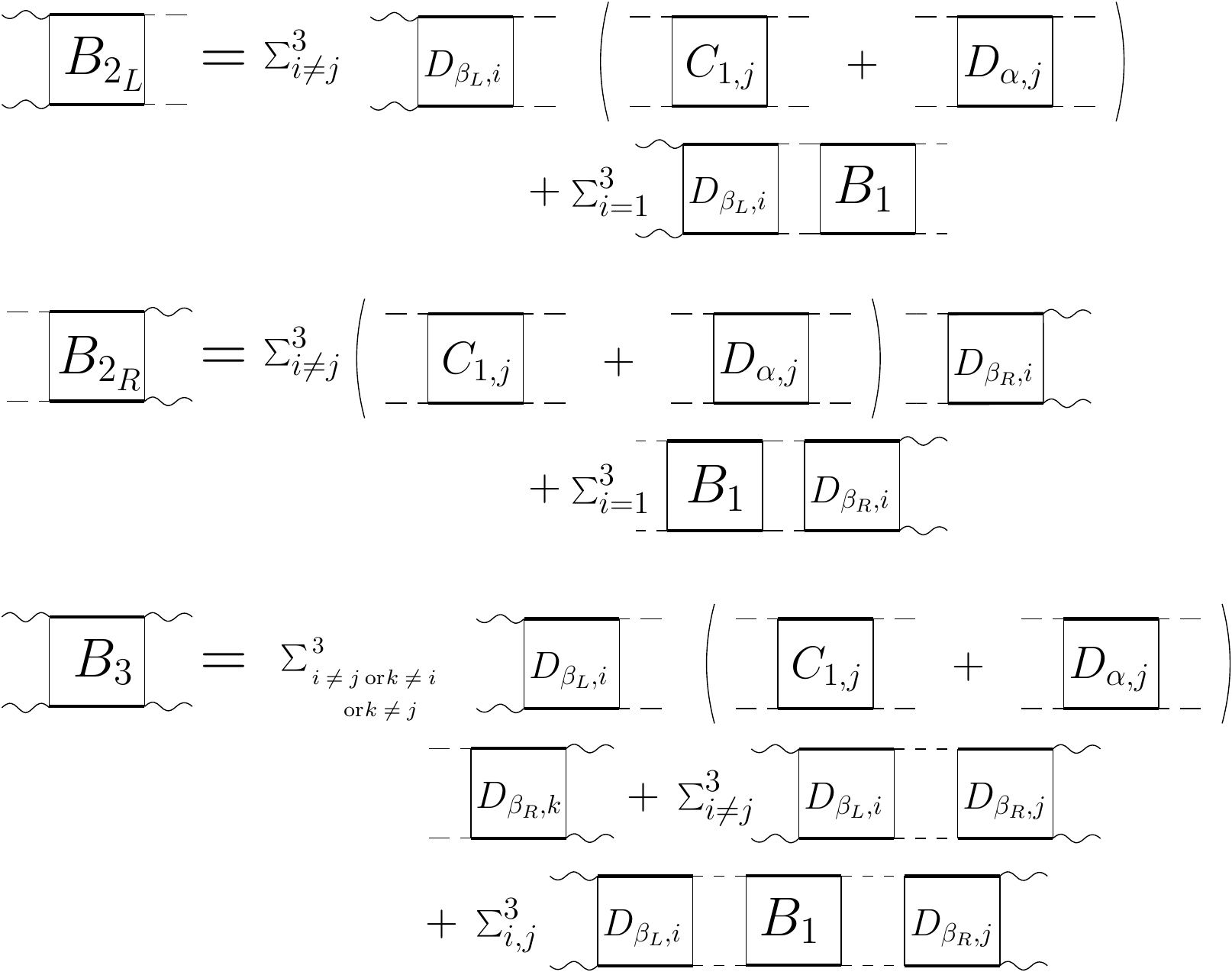}
\caption{The structure of the broken chains of family A.}
\label{broken_family_A}
\end{figure}

The generating functions of the broken chains of type 1 can be derived from the  expressions 
of the generating functions above
by noticing that every chain of type 1 that is not of a given color must be broken. 
When building chains, there are three possible colors of dipoles to insert. Multiplying  each generating functions of dipoles 
by a factor three,
and subtracting the chains of color $i$,  gives  the generating functions of the broken chains $B_1$.   This writes as:
\begin{equation}
B_1 = \frac{3 D_\alpha +9 D_{\beta_L} D_{\beta_R}}{1- (3 D_\alpha +9 D_{\beta_L} D_{\beta_R})} -3 D_\alpha - \sum_{i=1,2,3} C_{1,i}
\text{.}
\label{broke_chain_t1}  
\end{equation}
The expressions of the remaining types of broken chains follow then from Figure \ref{broken_family_A} and write:
\begin{equation}
\begin{split}
B_{2_L} &= 6 D_{\beta_L} ( C_{1} + D_{\alpha} ) + 3 D_{\beta_L} B_1 \text{,}\\
B_{2_R} &= 6( C_{1} + D_{\alpha} )  D_{\beta_R}+ 3 B_1  D_{\beta_R} \text{,} \\
B_{3} &= 24 D_{\beta_L} ( C_{1} + D_{\alpha} ) D_{\beta_R} + 6 D_{\beta_L} D_{\beta_R}  + 9 D_{\beta_L} B_1 D_{\beta_R}\text{.}
\end{split}
\end{equation}

After summing over the three colors 
in the last term of equation \eqref{broke_chain_t1},
and after some further algebra, one gets:
\begin{equation}
\begin{split}
&B_1 = \frac{6 D_{\beta_L} D_{\beta_R} + 18 (D_{\beta_L} D_{\beta_R})^2 + 6 D_\alpha^2 + 24 D_\alpha D_{\beta_L} D_{\beta_R}}{\big{(}1- (3 D_\alpha +9 D_{\beta_L} D_{\beta_R}) \big{)} \big{(}1- ( D_\alpha +D_{\beta_L} D_{\beta_R}) \big{)}} \text{,} \\
&B_{2_L} = D_{\beta_L}  \frac{24 D_{\beta_L} D_{\beta_R} + 6 D_\alpha  }{\big{(}1- (3 D_\alpha +9 D_{\beta_L} D_{\beta_R}) \big{)} \big{(}1- ( D_\alpha +D_{\beta_L} D_{\beta_R}) \big{)}} \text{,}\\
&B_{2_R} =  D_{\beta_R}  \frac{24 D_{\beta_L} D_{\beta_R} + 6 D_\alpha  }{\big{(}1- (3 D_\alpha +9 D_{\beta_L} D_{\beta_R}) \big{)} \big{(}1- ( D_\alpha + D_{\beta_L} D_{\beta_R}) \big{)}} \text{,}\\
&B_{3} = \frac{3 D_{\beta_L} D_{\beta_R} (2+ 6 D_{\beta_L} D_{\beta_R})}{\big{(}1- (3 D_\alpha +9 D_{\beta_L} D_{\beta_R}) \big{)} \big{(}1- ( D_\alpha + D_{\beta_L} D_{\beta_R}) \big{)}}  \text{.}
\end{split}
\label{eq_broken_fam_A}
\end{equation}

\begin{figure}[h!]
\centering
\includegraphics[scale=0.95]{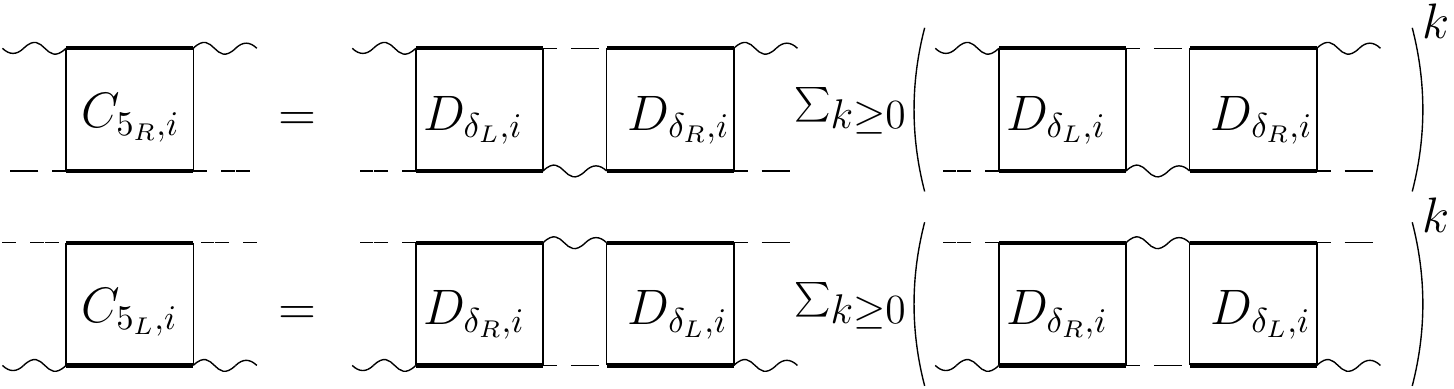}
\caption{Structure of the chains of type 5.}
\label{chain_type7_8}
\end{figure}

\paragraph{Family B:}
As above, the structure of the chains in family B is derived from their building blocks that are dipoles of type $\delta_L$ and $\delta_R$. If a dipole of type $\delta_L$ or resp. $\delta_R$ doesn't end a chain, it must be followed by a dipole of type $\delta_R$ or resp. $\delta_L$. All the possibilities for type $5$ chains are then given in Figure \ref{chain_type7_8}.

\begin{figure}[h!]
\centering
\includegraphics[scale=0.8]{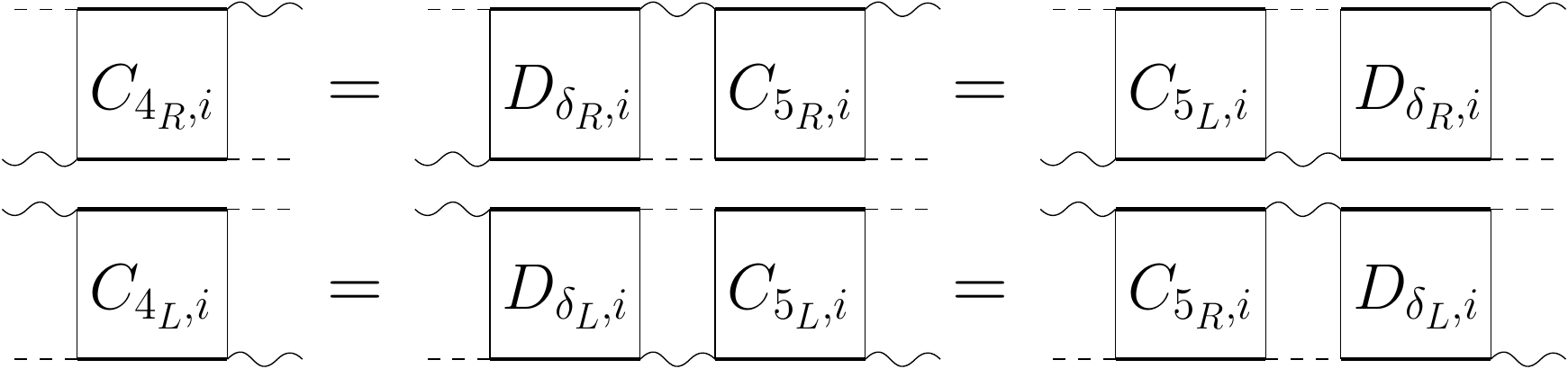}
\caption{Structure of the chains of type 4.}
\label{family_B}
\end{figure}

The structure of the others chains can then be deduced from the type $5$ chains and is shown in Figure \ref{family_B}.

The generating functions of family $B$ chains write:
\begin{equation}
\begin{split}
C_{5_L} = C_{5_R} &=  D_{\delta_L} D_{\delta_R} \sum_{k \geq 0} (D_{\delta_L} D_{\delta_R})^{k} = \frac{D_{\delta_L} D_{\delta_R}}{1- D_{\delta_L} D_{\delta_R}} \text{,}\\
C_{4_L} &= D_{\delta_L} C_{5_L} = \frac{D_{\delta_L}^2 D_{\delta_R}  }{1- D_{\delta_L} D_{\delta_R}} \text{,}\\
C_{4_R} &= D_{\delta_R} C_{5_R} = \frac{D_{\delta_L} D_{\delta_R}^2 }{1- D_{\delta_L} D_{\delta_R}} \text{.}
\end{split}
\end{equation}
Using a similar argument as in the case of family A chains, the broken chains of this family are found to be
\begin{equation}
\begin{split}
B_{5_L} &= B_{5_R} =  D_{\delta_L} D_{\delta_R} \frac{6 + 18 D_{\delta_L} D_{\delta_R}}{(1-D_{\delta_L} D_{\delta_R}) (1-9 D_{\delta_L} D_{\delta_R})} \text{,}\\
B_{4_L} &=   \frac{24 D_{\delta_L}^2 D_{\delta_R}}{(1-D_{\delta_L} D_{\delta_R}) (1-9 D_{\delta_L} D_{\delta_R})}\text{,}\\
B_{4_R} &=  \frac{24 D_{\delta_L} D_{\delta_R}^2}{(1-D_{\delta_L} D_{\delta_R}) (1-9 D_{\delta_L} D_{\delta_R})}  \text{.}\\
\end{split}
\end{equation}

The formulas obtained above for the generating functions of the various types of chains have therefore five different types of singular points:
\begin{itemize}
\item As any dipole has the same singular points as the melons, the chains also have the same singular points as the melons. 
\item For the unbroken and broken chains of family A, the points satisfying $D_\alpha + D_{\beta_L} D_{\beta_R} =1$ are singular points.
\item For the broken chains of family A, the points satisfying $3 D_\alpha + 9 D_{\beta_L} D_{\beta_R} =1$ are singular points.
\item For the colored and broken chains of family B, 
 the points satisfying
$D_{\delta_L} D_{\delta_R} =1$ are singular points.
\item For the broken chains of family B, 
 the points satisfying
$9 D_{\delta_L} D_{\delta_R} =1$ are singular points.     
\end{itemize}
Inserting the expression of the generating functions of the dipoles in the singularity conditions above, one has:
\begin{equation}
\begin{split}
& \frac{2}{3} (M_T(t)-1) + \frac{1}{9} (M_T(t)-1)^2 = 1 \text{,}\\
& 2 (M_T(t)-1) + (M_T(t)-1)^2 = 1 \text{,}\\
& \frac{1}{9} (M_T(t)-1)^2 = 1 \text{,}\\
& (M_T(t)-1)^2 = 1 \text{.}
\end{split}
\end{equation}
Each condition leads to a $2^{nd}$ order equation and the solutions of these equations are:
%\begin{equation}
%\begin{split}
$M_T = -2 \pm 3 \sqrt{2} \text{,}\ 
M_T = \pm \sqrt{2} \text{,}\ 
M_T = 4 \text{, } -2$ and resp.
$M_T = 2 \text{, } 0$.
%\end{split}
%\end{equation}
The points where $3 D_\alpha + 9 D_\beta^2 =1$ are therefore also points where the melons are singular. Recall that $M_T(t)$ was chosen to be an increasing function of $t$. This means that the singularity with the smallest $|M_T|$ is also the one with $t$ closest to zero. Inserting $M_T = 0$ in equation \eqref{Melonic_SD} gives $1=0$ and is therefore impossible. The dominant singularity is then the point $(M_{T,c},t_c) = (\sqrt{2},\frac{1}{8 \sqrt{2}+12})$ which is a singular point for the generating functions of melons and broken chains of family A.
 
 Notice also from equation \eqref{eq_broken_fam_A} that only the broken chains of type $2_L$,$2_R$, $3$ that are composed of a broken chain of type 1 are singular at the critical point. We denote these specific chains by a star such that $B_{2_L^*} =  3 D_{\beta_L} B_1$, $B_{2_R^*} =  3 D_{\beta_R} B_1$ and $B_{3^*} = 9 D_{\beta_L} B_1 D_{\beta_R} $.

Then, near the critical point, one gets:

\begin{equation}
\begin{split}
B_1 & \stackrel{t \rightarrow t_c}{\approx} \frac{3 D_\alpha +9 D_\beta^2}{1- (3 D_\alpha +9 D_\beta^2)} \stackrel{t \rightarrow t_c}{\sim} \frac{1}{2 M_{T,c} K \sqrt{1-\frac{t}{t_c}}} \text{,}\\
B_{2_L} &   \stackrel{t \rightarrow t_c}{\approx} B_{2_L^*} = 3 D_{\beta_L} B_1 \stackrel{t \rightarrow t_c}{\sim} \frac{M_{T,c}-1}{ M_{T,c} K \sqrt{1-\frac{t}{t_c}}} \text{,}\\
B_{2_R} &   \stackrel{t \rightarrow t_c}{\approx} B_{2_R^*} = 3 D_{\beta_R} B_1 \stackrel{t \rightarrow t_c}{\sim} \frac{M_{T,c}-1}{ M_{T,c} K \sqrt{1-\frac{t}{t_c}}} \text{,}\\
B_{3} & \stackrel{t \rightarrow t_c}{\approx}  B_{3^*} = 9 D_{\beta_L} D_{\beta_R} B_1 \stackrel{t \rightarrow t_c}{\sim} \frac{(M_{T,c}-1)^2}{ M_{T,c} K \sqrt{1-\frac{t}{t_c}}} \text{.}
\end{split}
\label{lim_B}
\end{equation}
%Hence, the most singular schemes close to criticality are the one ththat maximize the number of broken chains of family A. 

\subsection{Scheme decomposition}

%As both the length of a chain can be changed and a melon can be inserted/eliminated on any propagator without changing the degree, there is an infinite number of graphs at a given degree. The idea behind the scheme decomposition is to contract these infinities into a finite number of graphical objects (which are the schemes). 

%In the tetrahedric representation, 
As in \cite{Bonzom_2022} \cite{https://doi.org/10.48550/arxiv.1408.5725},
the scheme  $\mathcal{S}$
of a graph $\mathcal{G}$ is defined by eliminating 
from the respective graph 
any $2$-point melonic subgraph, and replacing each maximal chain by its corresponding chain-vertex and any dipole by its corresponding dipole-vertex. 

If a scheme is derived from a rooted graph, then the respective scheme naturally contains a rooted edge.

One can prove that the degree of the scheme $\omega(\mathcal{S})$ is identical to the one of the initial graph $\omega(\mathcal{G})$ (see again 
 \cite{Bonzom_2022} \cite{https://doi.org/10.48550/arxiv.1408.5725}).
 
Each scheme represents then a family of graphs and any graph can be derived from its corresponding scheme by replacing/inserting back the corresponding dipoles, chains and melons. 

\begin{figure}[h!]
\centering
\includegraphics[scale=1]{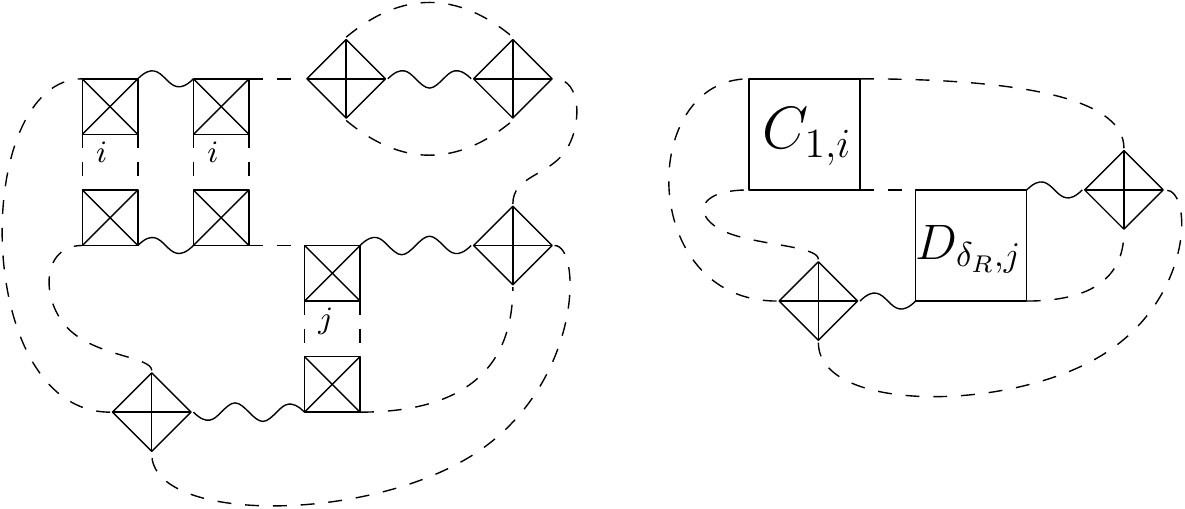}
\caption{Example of a graph (left) and its corresponding scheme (right).}
\label{example_scheme}
\end{figure}

Let us give an example of a scheme and a rooted Feynman graph corresponding to it in Fig. \ref{example_scheme}.

In \cite{Bonzom_2022} it was proven that the number of schemes at a given degree is finite in the $O(N)^3$-invariant model. In order to obtain this result, it was proven that both the number of dipole/chain-vertices and the number of tetrahedric vertices was bounded. 

The schemes of the prismatic model can be obtained by decorating the edge that links pairs of vertices of the schemes present in the tetrahedric model. There is also only a finite number of ways to perform such decoration on a scheme with a given number of dipole-vertices, chain-vertices, and tetrahedric vertices. This implies that the number of schemes at a given degree in the prismatic model must be finite. 

As the number of schemes at each degree is finite, we can use them to compact the infinite number of graphs at a fixed degree into a finite number of schemes. Thus, the sum over an infinite number of Feynman graphs %at a fixed degree $\omega$ 
in $G_2(t)$ can be written as a sum over a finite number of schemes. The $2-$point function then writes
\begin{equation}
G_2(t) = M_T(t)  + \sum_{\omega \geq 1} N^{-\omega} \sum P_{\mathcal{S},\omega}(M,C) \text{,}
\end{equation}
where $P_{\mathcal{S},\omega}(M,C)$ is a polynomial of the melons generating function and the chains generating functions.
The sum in the formula above is performed on all the schemes of degree $\omega$. 
%As $\mathbb{S}_\omega$ contains a finite number of elements, the singularities of $G_2(t)$ may only arise from the generating functions $M_T(t)$ , $D(t)$, $C(t)$ and $B(t)$. 
The dominant singularity of $G_2(t)$ is then the singular point $(M_{T,c},t_c) = (\sqrt{2},\frac{1}{8 \sqrt{2}+12})$ of the broken chains of type $1,2_L^*,2_R^*$ and $3^*$.

\begin{figure}[h!]
\centering
\includegraphics[scale=1]{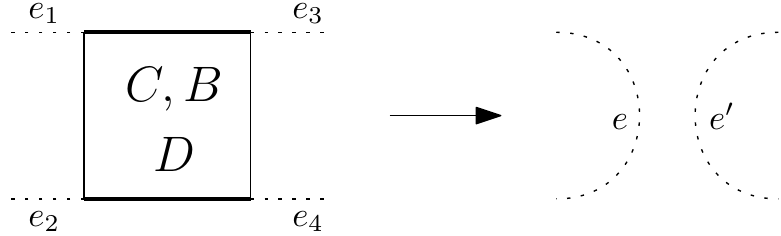}
\caption{The process of dipole removal.}
\label{dipole_removal}
\end{figure}

\paragraph{Dipole/chain-vertex removal}   When studying the combinatorics schemes, a standard operation is to remove dipole/chain-vertices and to reconnect the half-edges together on each sides of the vertices as in Figure \ref{dipole_removal}.

A dipole/chain-vertex is said to be separating if its removal disconnects the scheme $\mathcal{S}$. One can prove that the removal of a separating dipole/chain-vertex leads to a distribution of  the degree between the two connected components (see again
\cite{Bonzom_2022} and \cite{https://doi.org/10.48550/arxiv.1408.5725}).
Thus, removing a separating dipole/chain-vertex 
leaves the degree invariant. Moreover, one has
\begin{equation}
\omega(\mathcal{S}) = \omega(\mathcal{S}_1) +  \omega(\mathcal{S}_2) \text{,}
\label{separating_dipole}
\end{equation}
where $\mathcal{S}_1$ and $\mathcal{S}_2$ are the two connected components resulting from the removal.

Let us denote by $\mathcal{S}'$ the scheme obtained after a dipole removal in $\mathcal{S}$. By carefully counting the number of faces and vertices,  one can prove 
(see again
\cite{Bonzom_2022} and \cite{https://doi.org/10.48550/arxiv.1408.5725})
that removing a non separating dipole or chain of color $i$ changes the degree as 
\begin{equation}
\omega(\mathcal{S}) -1 \geq \omega (\mathcal{S}') \geq  \omega(\mathcal{S}) -3 \text{.}
\label{degree_vary}
\end{equation}
Analogously, removing a non separating broken chain gives $\omega (\mathcal{S}') = \omega(\mathcal{S}) -3$

Following \cite{Bonzom_2022}, let us give here some details on the proof of the  \eqref{separating_dipole} and \eqref{degree_vary}. Recall that changing the length of a chain doesn't change the degree of a scheme. Therefore, when studying the face structure of chains, one can consider the minimal case that are chains of length $2$.

\begin{figure}
    \centering
    \includegraphics[scale=0.7]{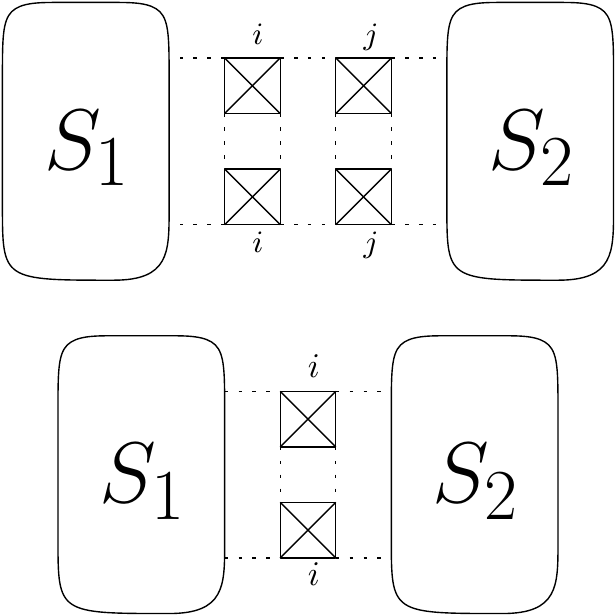}
    \caption{Structure of a scheme possessing a separating dipole/chain-vertex.}
    \label{fig:separating}
\end{figure}

\begin{figure}
    \centering
    \includegraphics[scale=0.7]{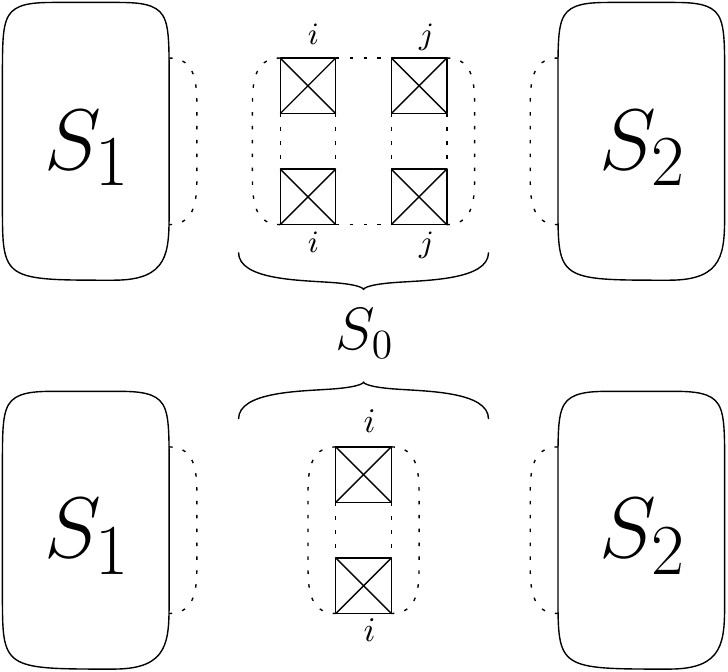}
    \caption{The three components that can be obtained from Figure \ref{fig:separating}.}
    \label{fig:separating_cutted}
\end{figure}

\subparagraph{Separating dipole/chain-vertex.} If a scheme $\mathcal{S}$ possesses a separating dipole/chain-vertex, it is of one of the forms given in Figure \ref{fig:separating}, where the edges are either $\chi$ or $T$ propagators. The scheme has then two $2-$edge cuts at each side of the dipole/chain-vertex. Performing the two cuts leads to a new scheme $\mathcal{S}'$ with $3$ disconnected components, denoted by $\mathcal{S}_0$, $\mathcal{S}_1$, $\mathcal{S}_2$, as given in Figure \ref{fig:separating_cutted}. The number of faces of $\mathcal{S}'$ is $f_{\mathcal{S}'}= f_{\mathcal{S}_1} + f_{\mathcal{S}_2} + f_{\mathcal{S}_0} =  f_{\mathcal{S}}+6 $ and the number of vertices $n_{\mathcal{S}'}$ is equal to $n_\mathcal{S}$. The degree $\omega(\mathcal{S})$ then writes:
\begin{equation*}
    \omega(\mathcal{S}) = 3 + n_{\mathcal{S}} - f_{\mathcal{S}} =  3 + n_{\mathcal{S}_1} + n_{\mathcal{S}_2}+ n_{\mathcal{S}_0} - f_{\mathcal{S}_1} - f_{\mathcal{S}_2} - f_{\mathcal{S}_0} + 6  \text{ .}
\end{equation*}
This can be rewritten as 
\begin{equation*}
    \omega(\mathcal{S}) = \omega(\mathcal{S}_1) + \omega(\mathcal{S}_2) + \omega(\mathcal{S}_0) \text{ ,}
\end{equation*}
and since the component $\mathcal{S}_0$ is a degree $0$ component it gives:
\begin{equation*}
    \omega(\mathcal{S}) = \omega(\mathcal{S}_1) + \omega(\mathcal{S}_2) \text{ .}
\end{equation*}

\begin{figure}
    \centering
    \includegraphics[scale=1]{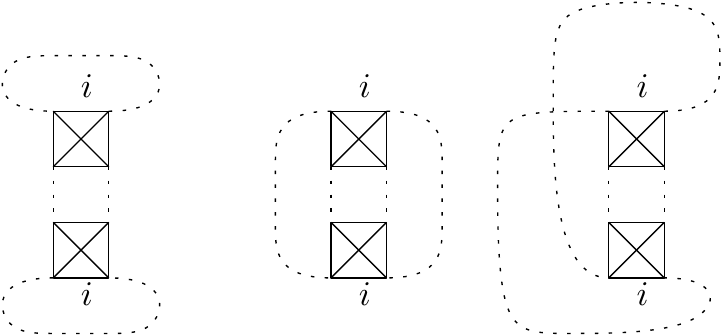}
    \caption{The three possible structures of the faces of color $i$ adjacent to a color $i$ dipole.}
    \label{fig:structure}
\end{figure}

\subparagraph{Non-separating dipole vertex.} In the case of a color $i$ dipole only the faces of color $i$ are affected by the removal. There are either $1$ or $2$ faces of color $i$ adjacent to such dipoles as shown in Figure \ref{fig:structure}. Removing the dipole reduces the number of vertices by $2$, and the number of faces by either $0$, $1$ or $2$. This implies that, for a non-separating dipole removal, the variation of the degree $\Delta \omega$ is
\begin{equation*}
    -3 \leq \Delta \omega \leq -1 \text{ .}
\end{equation*}

\subparagraph{Non-separating chain vertex.} When removing a chain vertex, one needs to separate the broken case to the colored one. If the chain is broken, the chains incident to the vertex are not affected by the removal. One can then count the variation of the number of internal faces  and of vertices to obtain the variation of the degree. Doing so on the case of a chain of length $2$ gives
\begin{equation*}
    \Delta \omega = 3\ .
\end{equation*}

If the chain is not broken, the analysis is the same as the case of a non separating dipole removal, leading to  
\begin{equation*}
    -3 \leq \Delta \omega \leq -1 \text{ .}
\end{equation*}
\paragraph{Skeleton graphs}

When identifying the dominant schemes, a key role is played by skeleton graphs (see again \cite{Bonzom_2022}). These graphs are defined as follows. 

Consider a scheme $\mathcal{S}$, its skeleton graph $\mathcal{I} (\mathcal{S})$ is obtained by removing all its broken chain-vertices of type $1$ and by adding arrows (labeled by $B_1$) between the edges formed in this way. Notice that the definition given here is slightly different than the one given in \cite{Bonzom_2022}. 

Again, if the skeleton graph is built from a rooted scheme, it will also contain a root.

\begin{figure}[h!]
\centering
\includegraphics[scale=1]{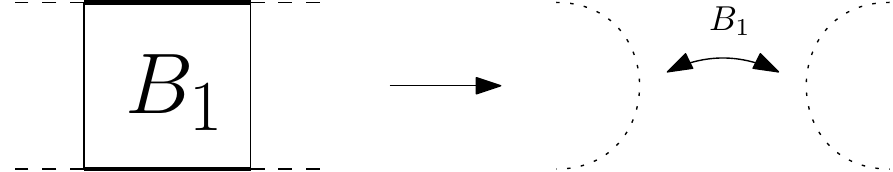}
\caption{The dipole removal in the construction of the skeleton graph.}
\label{dipole_removal_skeleton}
\end{figure}

\begin{figure}[h!]
\centering
\includegraphics[scale=0.7]{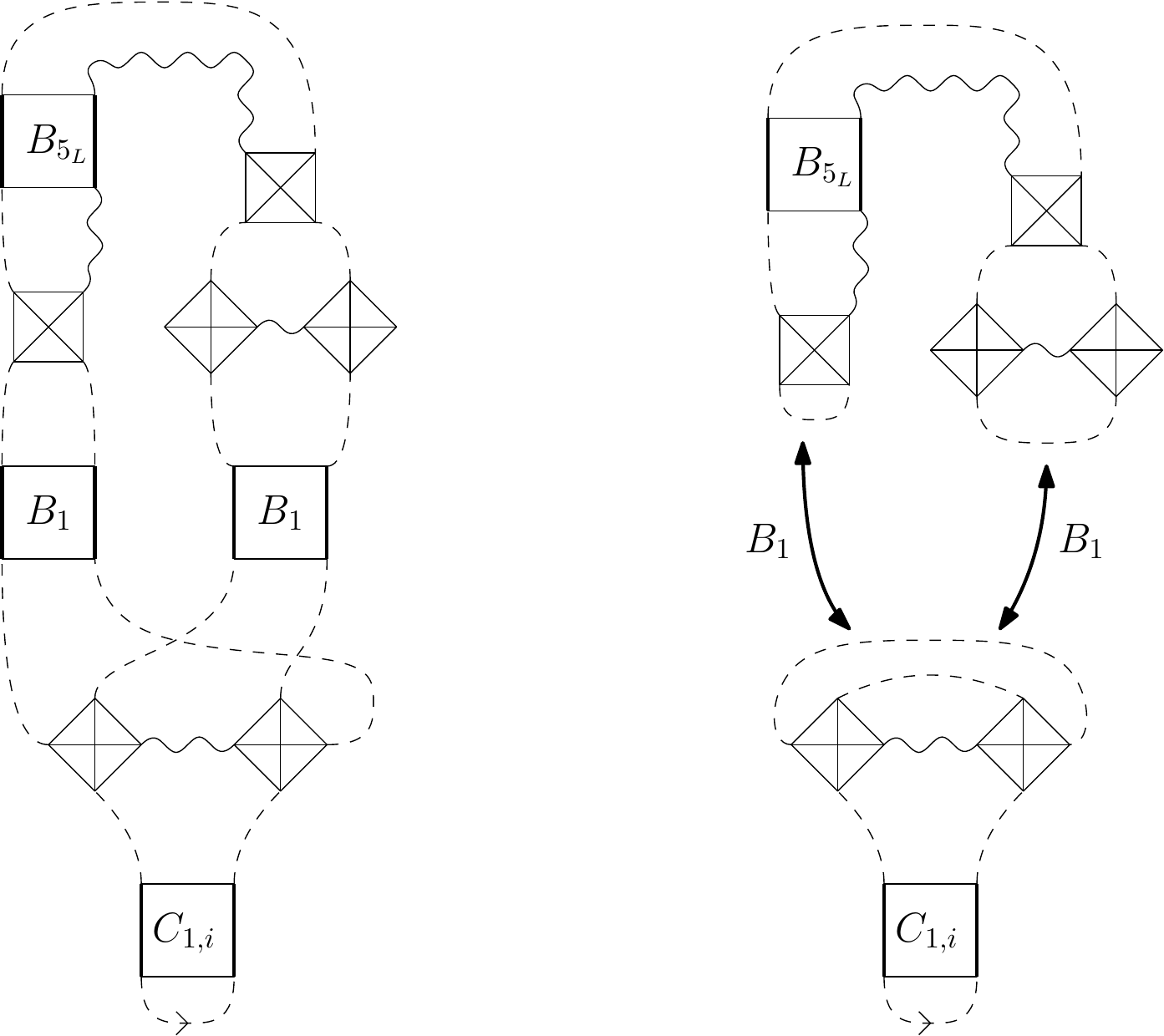}
\caption{An example of a scheme and its skeleton graph.}
\label{skeleton_graph}
\end{figure}

An example of such a dipole-vertex removal is given in Figure \ref{dipole_removal_skeleton}. In a skeleton graph, the arrows play the role of the edges whereas the disconnected components $\mathcal{S}_i$, resulting from the removals, play the role of the vertices. However, these edges and vertices don't contribute to the degree as in a Feynman graph. An example of a scheme and its skeleton graph is given in Figure \ref{skeleton_graph}.

Skeleton graphs have several interesting properties (see again \cite{Bonzom_2022}): 
%that allow them to be studied more easily than the schemes. By using them, results on the schemes can be derived. The $3$ main properties are:
\begin{itemize}
\item If one of the components $S_i$ has degree $0$, then this component has a valency greater or equal to $3$.

 This can be proven by noticing that a degree $0$ graph with valency $1$ must be a melonic $2$-point function and one with valency $2$ must be a dipole or chain. None of these situations can occur for a skeleton graph.
 
\item Let $\mathcal{S}$ be a scheme with $q$ non separating chain-vertices of type $1$ and let $\mathcal{S}'$ the scheme obtained be removing these
$q$ chain-vertices. The skeleton graph $\mathcal{I}(\mathcal{S}')$ is a tree and its degree satisfies the inequality: $\omega ( \mathcal{I}(\mathcal{S})) \leq \omega (S) - q$.

%A tree is a graph with no cycle. 
This comes from the fact that, if a scheme has only separating dipole/chain-vertices no cycle can appear in its skeleton graph. The bound on $\omega ( \mathcal{I}(\mathcal{S}))$ is derived by using equation \eqref{degree_vary} and by recalling that the removal of a separating dipole/chain-vertex doesn't change the degree.

\item If $\mathcal{I}(\mathcal{S})$ is a tree, its degree is equal to the one of its scheme and is given by the sum of the degree of its components
\begin{equation}
\omega (\mathcal{I}(\mathcal{S})) = \sum_i \omega ( \mathcal{S}_i) \text{.}
\end{equation}
This comes from the fact that, 
if $\mathcal{I}(\mathcal{S})$ is a tree, all its dipole/chain-vertices are separating. This is then a direct consequence of equation \eqref{separating_dipole}.
\end{itemize}

%Trees are very well-known recursive structures that have been studied extensively in the past. 
%As the skeleton graph of any scheme that does not contain any non separating dipole/chain vertex is a tree, their properties can be used along with these skeletons graphs. 

\label{Skeleton}

\section{Next-to-leading order graphs}
\label{NLO}

In this section we  explicitly identify the NLO Feynman graphs  of the prismatic model. Recall from equation \eqref{degree_eq} that the degree of the prismatic model is a non-negative integer. Hence, if $\mathcal{G}$ is an NLO Feynman graph, one has: $\omega (\mathcal{G})=1$.

%As both $n_p$ and $f_\mathcal{G}$ are integers one has $\omega \in \mathbb{Z}_+$. This implies that the NLO graphs have degree 1. 

%In order to study these graphs, we work in the tetrahedric representation.
The graphs of degree one for the 
tetrahedric model 
have been identified in \cite{Nador2019}. 
Using this analysis, one can derive the corresponding NLO Feynman graphs of the model considered here. This is done using the following strategy. 
We consider all the graphs of degree one 
identified in \cite{Nador2019}, and investigate all the ways to decorate each graph with a $\chi$ propagator per pair of vertices.
%Finally, one needs to discard all the graphs where this consistent decoration is not possible. 
Finally, by contracting the intermediate field propagator $\chi$ and discarding the potential redundancies, all the graphs of degree one in the prismatic representation are found. 

In the tetrahedric representation, the NLO graphs can be separated into three classes (see again \cite{Nador2019}):
\begin{enumerate}
    \item 2PI, dipole-free Feynman graphs
    \item 2PI Feynman graphs with dipoles
    \item 2PR Feynman graphs
\end{enumerate}

In the sequel, we give the schemes of the graphs in the tetrahedric representation and give some examples of
Feynman graphs in both the tetrahedric and the prismatic representations.
%their prismatic equivalent. 
Recall that the number of tetrahedric vertices replacing a chain-vertex is arbitrary, and the same holds for
the corresponding number prismatic vertices (when using the prismatic representation of the model).
%\footnote{Recall that changing the length of a chain doesn't change the degree of a graph. Any number of prism can then be inserted in the examples if it is done such that the pattern is preserved.}.

\subsection{Dipole-free 2 particle irreducible graphs }

\begin{figure}[h!]
\centering
\includegraphics[scale=0.35]{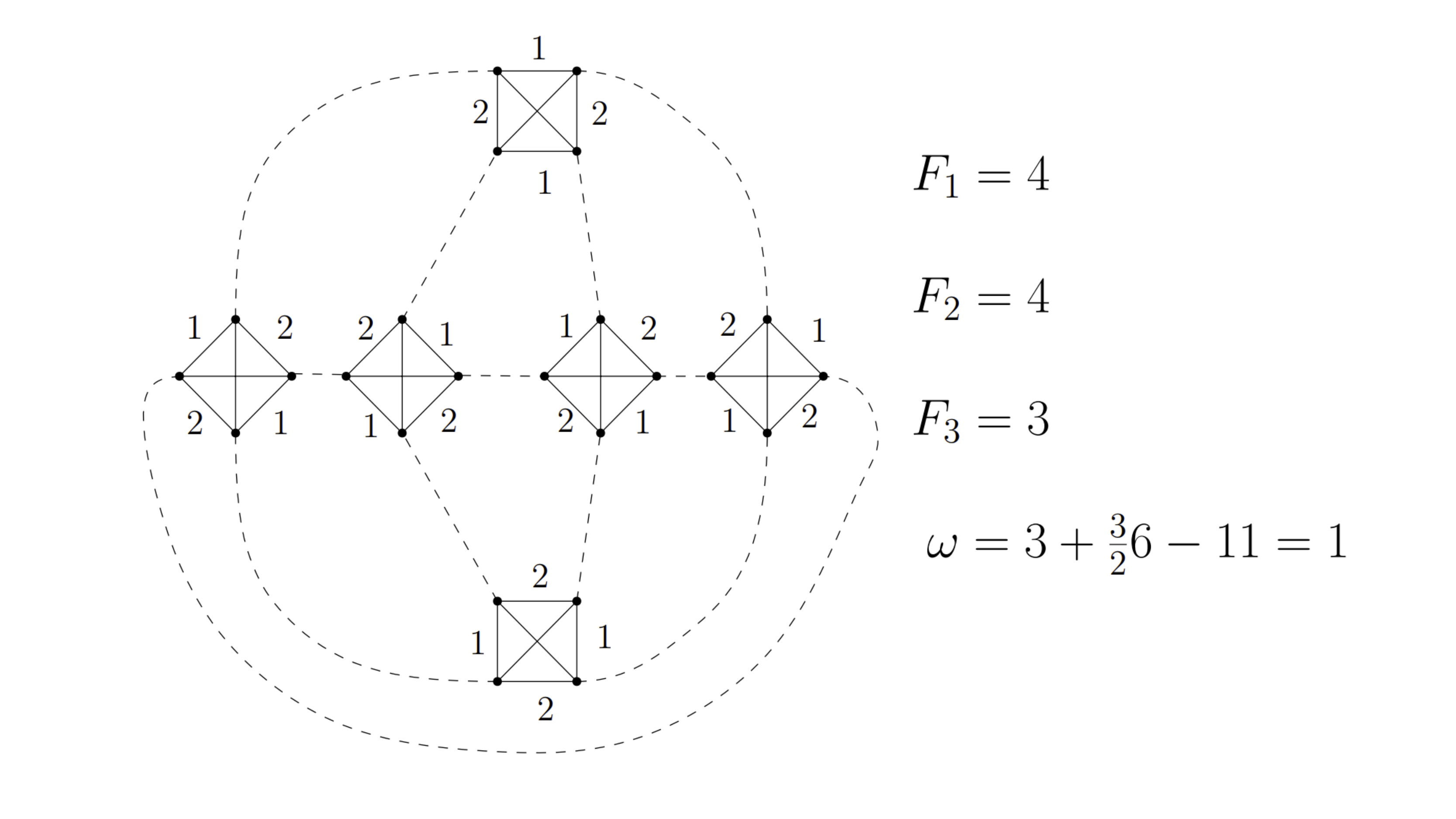}
\caption{The unique 2PI dipole free scheme of degree one in the tetrahedric model.}
\label{2pi_melonfree.pdf}
\end{figure}

It was derived in \cite{Nador2019} that there is a unique dipole free scheme of degree one in the tetrahedric model.  This scheme is given in Figure \eqref{2pi_melonfree.pdf}.

Note that, since this scheme has no dipole-vertices and no chain-vertices, one can obtain the corresponding graphs by the usual melonic insertions on any of its edges.

\begin{figure}[h!]
\centering
\includegraphics[scale=0.7]{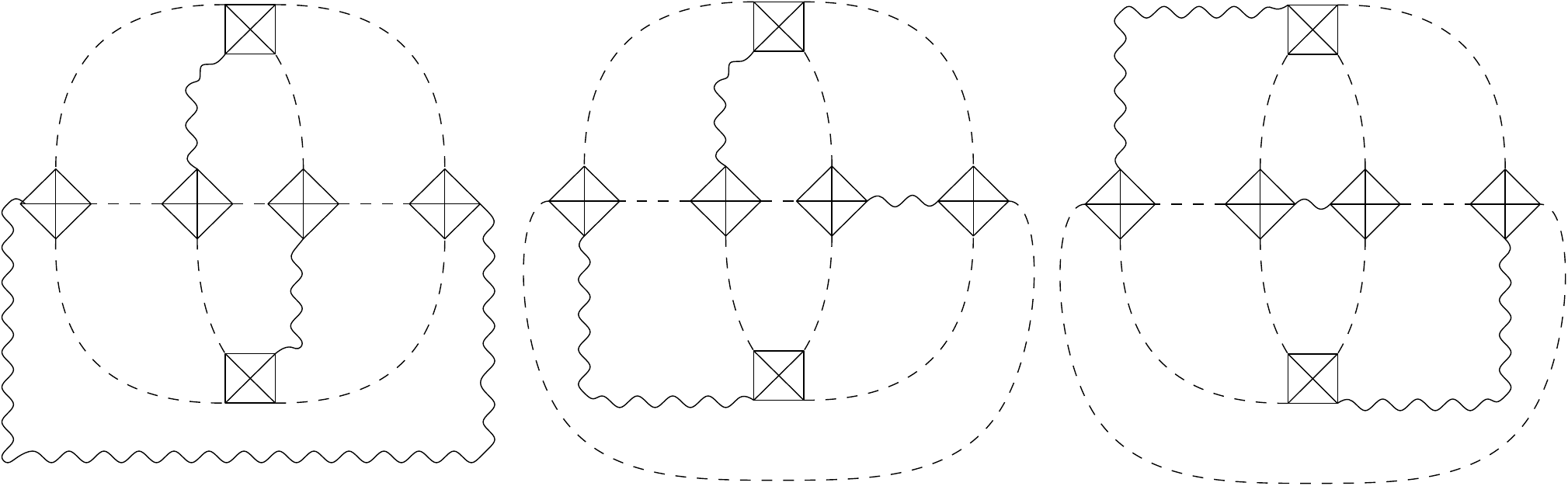}
\caption{The different ways to place the $\chi$ propagators on the NLO graph of Fig. \ref{2pi_melonfree.pdf}.}
\label{decorated_2pi_melon_free}
\end{figure}

There are then 3 independent ways to place the $\chi$ propagator on this NLO graph, see Fig. \ref{decorated_2pi_melon_free}.

\begin{figure}[h!]
\centering
\includegraphics[scale=1]{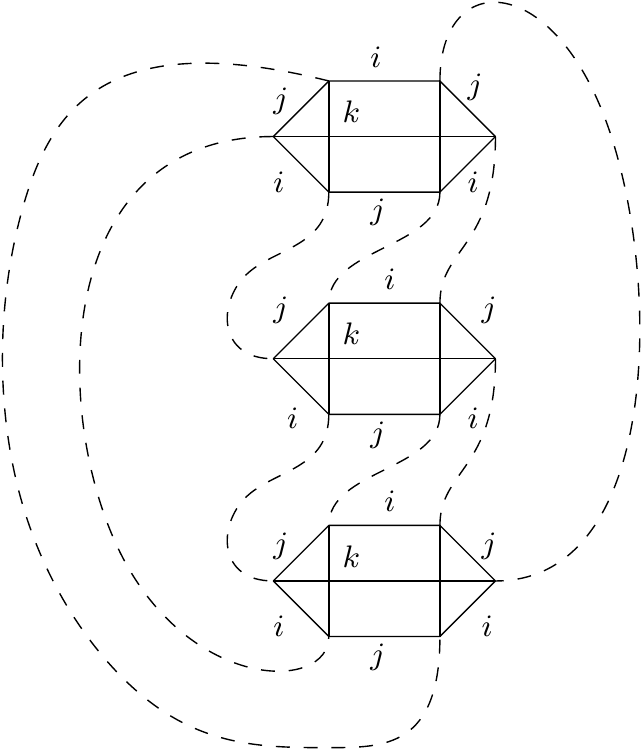}
\caption{The only graph found by contracting the $\chi$ edges of the graphs in Figure \ref{2pi_melonfree.pdf}.}
%{The 2PI dipole-free NLO graph, in the prismatic representation.}
%{The only graph found by contracting the $\chi$ lines introduced in Figure \ref{2pi_melonfree.pdf}.}
\label{graph_2_pi_prism}
\end{figure}

However, when contracting the $\chi$ propagators in order to obtain the prismatic representation NLO graphs, one can show that the three graphs of 
Fig.  \ref{decorated_2pi_melon_free}
%are equivalent and 
lead to the unique graph of Fig. \ref{graph_2_pi_prism}.

The NLO graphs in the prismatic representation are thus obtained 
{\it via} the two prismatic melonic moves (see above) performed 
on the graph of Fig. \ref{graph_2_pi_prism}.

\subsection{2PI graphs with dipoles}

\begin{figure}[h!]
\centering
\includegraphics[scale=1]{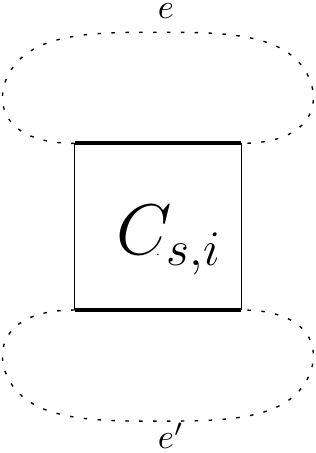}
\caption{The explicit schemes of the 2PI NLO graphs with dipoles.}
\label{chains_2pi_degree_1}
\end{figure}

\begin{figure}[h!]
\centering
\includegraphics[scale=1]{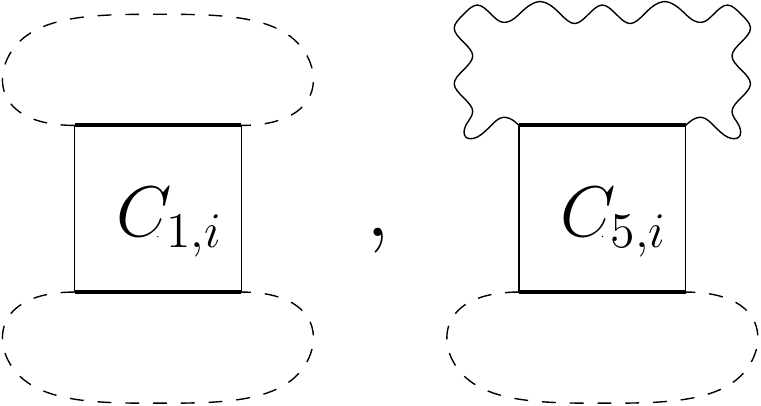}
\caption{The explicit schemes of the 2PI NLO graphs with dipoles.}
\label{chains_2pi_degree_12}
\end{figure}

The schemes of the 2PI graphs with dipoles in the tetrahedric representation are given in Figure \ref{chains_2pi_degree_1}. 
The edges $e$ and $e'$ can be of type $T$ or $\chi$, implying that $s=1$ or $5$ (the case $s=4$ beeing equivalent to $s=1$).  
The explicit schemes, with the corresponding $T$ or $\chi$ propagators, are given on the right side of Fig. \ref{chains_2pi_degree_12}.

%The chain vertex can be of type $t= 4,7$ and $8$, implying that the edges $e$ and $e'$ can be either a $T$ or $\chi$ propagator. 

\begin{figure}[h!]
\centering
\includegraphics[scale=0.7]{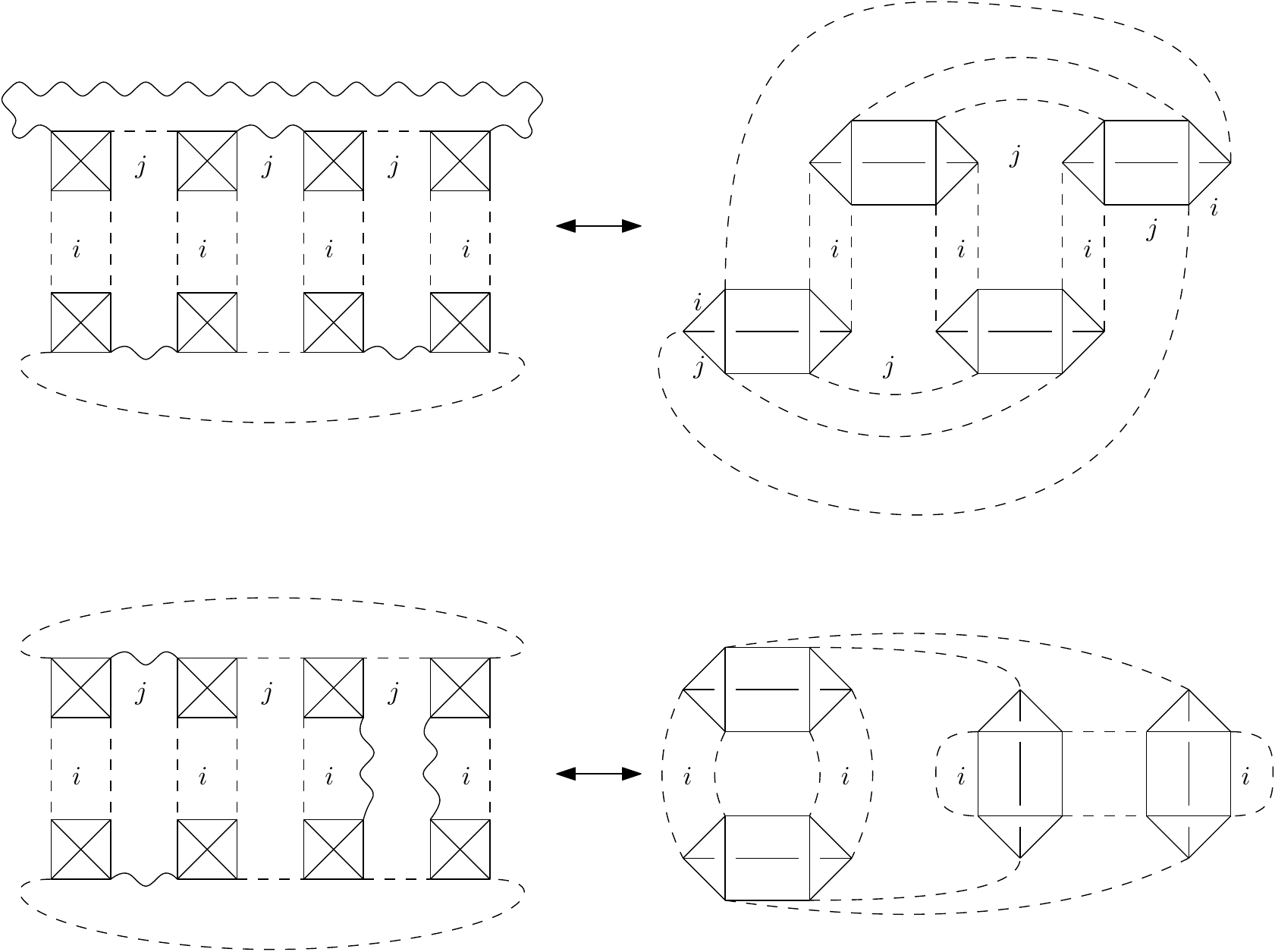}
\caption{Examples of NLO, 2PI graphs with dipoles in the tetrahedric (left) and prismatic (right) representation.}
\label{graph_2_pi_dipole_prism}
\end{figure}

Let us end this
subsection by giving some explicit examples of NLO 2PI graphs with dipoles (where we have replaced the chain-vertices by chains of length four), and their 
corresponding graphs in the prismatic representation - see Figure \ref{graph_2_pi_dipole_prism}.

\subsection{2 particle reducible NLO graphs}

\begin{figure}[h!]
\centering
\includegraphics[scale=0.6]{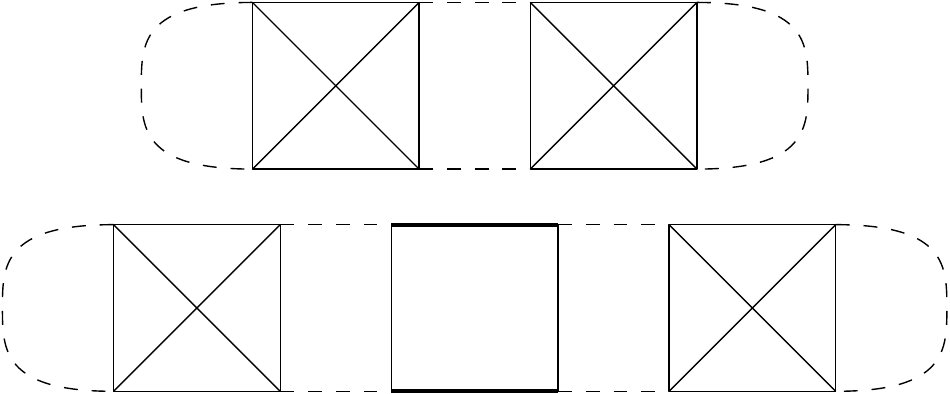}
\caption{The schemes of the NLO 2PR graphs in the single field tetrahedric model. }
\label{2_pr_single_degree_1}
\end{figure}

Recall from
\cite{Nador2019}
that schemes of the 2PR graphs in the tetrahedric model are given in Fig. \ref{2_pr_single_degree_1}.

\begin{figure}[h!]
\centering
\includegraphics[scale=0.6]{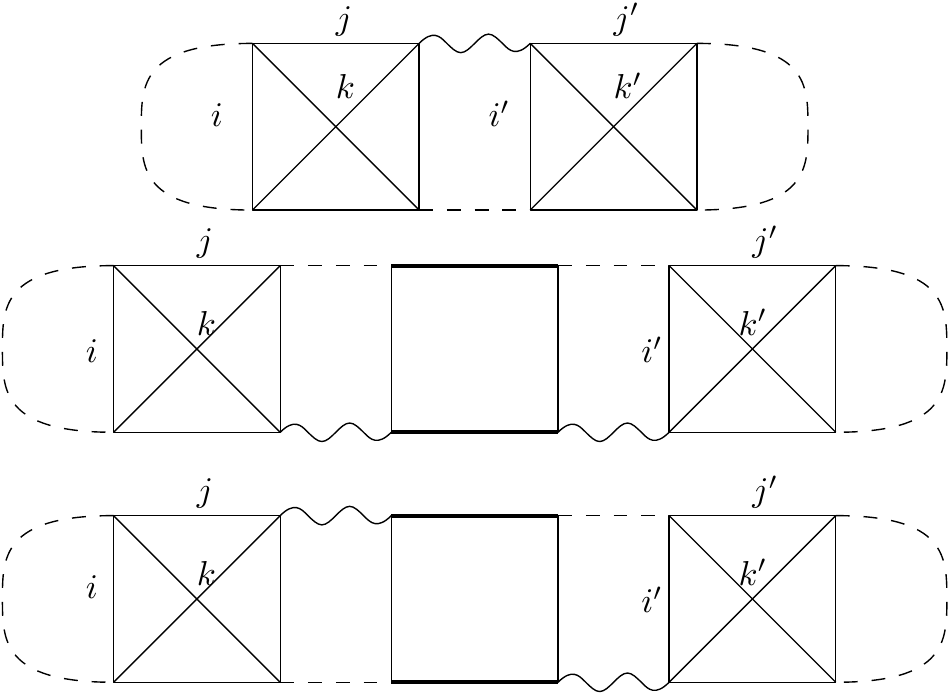}
\caption{The schemes of the NLO 2PR graphs. }
\label{2_pr_degree_1}
\end{figure}

When placing the $\chi$ propagators, the corresponding schemes
are given in Fig. \ref{2_pr_degree_1}. The empty boxes in the figures can either be a dipole-vertex, a broken chain-vertex or a colored chain-vertex.

\begin{figure}[h!]
\centering
\includegraphics[scale=0.9]{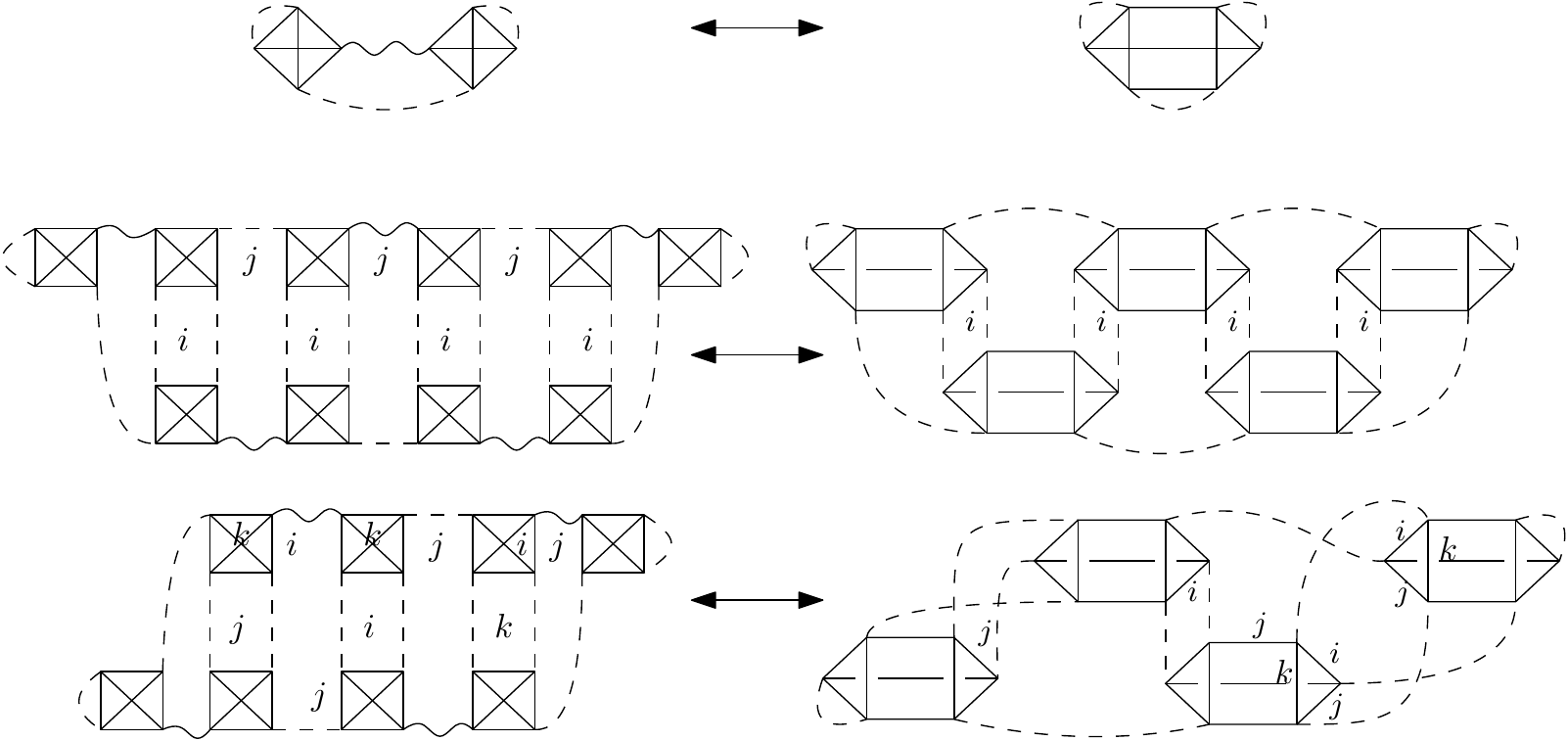}
\caption{Examples of the degree 1, 2PR graphs in the tetrahedric (left) and the prismatic (right) representation.}
\label{2_pr_degree_1_prism}
\end{figure}

Let us end this section by giving some examples of 
NLO 2PR graphs and the corresponding graph in the prismatic representation, see Figure \ref{2_pr_degree_1_prism}. The first graph in the figure is the first graph of Fig. \ref{2_pr_degree_1}. 
The second graph is obtained from the second scheme of Fig. \ref{2_pr_degree_1}, where we have replaced the empty box by a chain of length four and color $i$. Finally, the third graph 
is obtained from the second scheme of Fig. \ref{2_pr_degree_1}, where we have replaced the empty box by a broken chain of length four.
%the corresponding graphs are given in .

\section{Double scaling limit of the 2-point function}
\label{sec_DS}
Let us recall that since we are aiming to compute $G_2(t,N)$, we are working with rooted objects. Moreover, the dominant singularity of $G_2(t,N)$ is the point $(M_{T,c},t_c) =$ \\ $(\sqrt{2},\frac{1}{8 \sqrt{2}+12})$ where the broken chains of type $1,2_L^*, 2_R^*$ and $3^*$ are singular. The dominant schemes are then the ones that maximize the number of these chains.

%A canonical method in the study of the schemes consists of:
The general strategy used in this section is the following:
\begin{itemize}
\item  we remove all the non separating dipole/chain-vertices of a scheme $\mathcal{S}$; we denote by $\mathcal{S}'$ the resulting scheme.
\item we derive different bounds on the number of components of $\mathcal{S}'$ %by using its skeleton graph and the properties of trees;
\item finally, we use these bounds 
to obtain appropriate bounds on the sum of number of broken chain-vertices
of type $1$, $2_L^*$, $2_R^*$ and $3^*$
of
the original scheme $\mathcal{S}$.
%by using equation \eqref{degree_vary}. 
\end{itemize}

\begin{figure}[h!]
\centering
\includegraphics[scale=0.9]{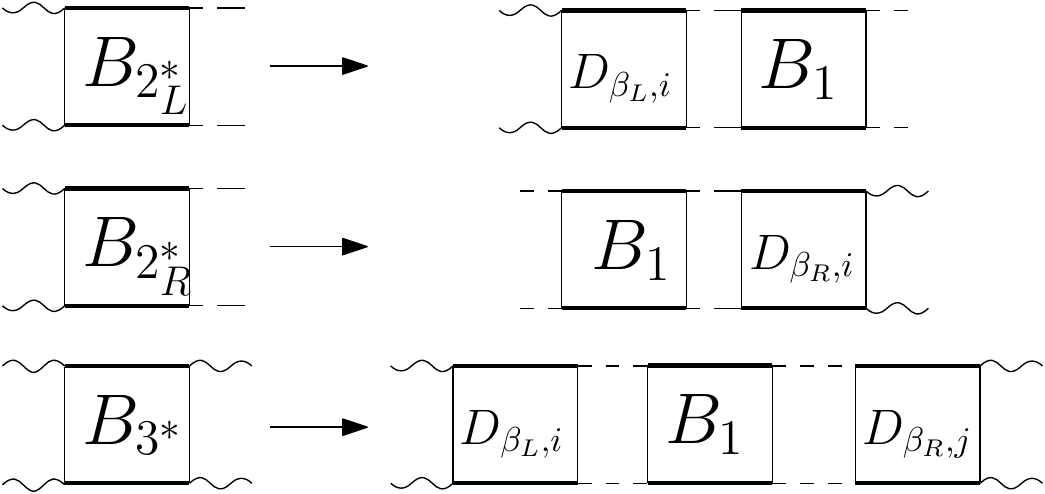}
\caption{Splitting of the chains of type $2^*,3^*$.}
\label{splitting_broken}
\end{figure}

Thus, let us consider a scheme $\mathcal{S}$ with $b$ the sum of the number of broken chains of type $1,2_L^*,2_R^*,3^*$. Note that these broken chains can be separating or non-separating, and we denote by 
$p$ and resp. $q$ the number of separating and resp. non-separating broken chains:
\begin{equation}
    b=p+q.
\end{equation}
We need to find a bound on $b$. In order to do so, as explained above, we remove all the $q$ non-separating broken chain-vertices and denote by $\mathcal{S}'$ the resulting scheme. The degree of $\mathcal{S}'$ is then $\omega(\mathcal{S}') = \omega(\mathcal{S}) -3 q$, and the resulting scheme has $p$ separating broken chains of type $1,2_L^*,2_R^*,3^*$. 
We then split every broken chain-vertex of type $2_L^*$ resp. $2_R^*$ into a dipole-vertex (of any color) of type $\beta_L$ resp. $\beta_R$ and a broken chain-vertex of type $1$. Similarly, we split every broken chains-vertex of type $3^*$ into two dipoles vertices of type $\beta$ and a broken chain-vertex of type 1. The graphical representation of this splitting procedure is given in Figure \ref{splitting_broken}. After this splitting, the scheme $\mathcal{S'}$ has $p$ separating broken chains-vertices of type $1$.

Consider the skeleton graphs $\mathcal{I} (\mathcal{S}')$ of $\mathcal{S}'$. As $\mathcal{S}'$ had only separating broken chains, $\mathcal{I} (\mathcal{S}')$ is a tree and its degree is given by the sum of the degree of its components.  
%Since the removal of a chain of family A doesn't introduce mixed lines in the scheme, 
Recall that the smallest positive degree possible for the components of $\mathcal{I} (\mathcal{S}')$ is one.

Since we are studying the $2-$point function, the skeleton graph $\mathcal{I} (\mathcal{S}')$ possesses a root and has 3 types of connected components:
\begin{itemize}
    \item The rooted component, it is unique.
    \item Non rooted components of positive degree, we denote their number by $N_+$.
    \item Non-rooted components of degree zero. Such components are said to be {\it tracked} if they are connected to a non separating broken chain-vertex of type $1$ in $\mathcal{S}$. Otherwise the components are said to be non-tracked. The number of non rooted degree $0$ components is denoted by $N_0 = N_{0,t} + N_{0,nt}$ where $N_{0,t}$ resp. $N_{0,nt}$ is the number of tracked resp. non-tracked non-rooted degree $0$ components.
\end{itemize}
Therefore, the number of components $N_c$ of the skeleton graph reads 
\begin{equation}
N_c= N_{0,nt} + N_{0,t} + N_{+} + 1. 
\end{equation}
Recall that in the skeleton graph $\mathcal{I} (\mathcal{S})$, a degree $0$ component has a valency at least equal to three. This implies that in $\mathcal{I}(\mathcal{S}')$ the valency of a tracked degree component is at least equal to one since it is at most connected to two separating broken chain-vertices of type $1$ in $\mathcal{S}$.

%The bounds on their numbers are given by: 
Following \cite{Bonzom_2022}, one can prove the following bounds:
\begin{equation}
%\label{1}
N_{0,t} \leq 2 q \text{,}
\label{positive}
\end{equation}
\begin{equation}
\label{2}
N_{+} \leq \omega(\mathcal{S}') \text{,}
\end{equation}
\begin{equation}
\label{3}
2 p \geq 3 N_{0,nt} + N_{0,t} + N_{+} + 1 \text{,}
\end{equation}
\begin{equation}
p =  N_{0,nt} + N_{0,t} + N_{+} \text{.}
\end{equation}
%Substituting the different bounds and using $\omega(\mathcal{S}') = \omega(\mathcal{S}) -3 q$ gives
After some algebra, one gets the following inequality
\begin{equation*}
p \leq 2 \omega( \mathcal{S}') + 4 q -1,
\end{equation*}
using $\omega ( \mathcal{S}') = \omega ( \mathcal{S}) - 3q $ one gets
\begin{equation}
p \leq 2 \omega( \mathcal{S}) - 2 q -1,
\end{equation}
%This implies that
which further leads to:
\begin{equation}
b \leq 2 \omega( \mathcal{S}) - q -1 \text{.}
\end{equation}
Hence the maximal value for $b$ is $2 \omega( \mathcal{S}) - 1$ (when 
%number of broken chains of type $1$,$2^*$,$3^*$ in a scheme of degree $\omega( \mathcal{S})$ is $2 \omega( \mathcal{S})-1$. This occurs when 
$q=0$). One can check that when $b$ reaches the upper bound, 
the inequalities 
\eqref{positive}, \eqref{2} and \eqref{3}
turn into equalities.
This implies that $N_{0,t} = 0$, $N_{+} = \omega (\mathcal{S})$ and that the valency of all degree $0$ components is $3$, while the one of the positive degree components is $1$.  

All positive components in $\mathcal{I} (\mathcal{S}')$ are of degree one, hence there are two possibilities. The components connected to a broken chain of type $1$, $2_L^*$ or $2_R^*$ (on the $T$ propagator side) before the removal are given by the NLO schemes described in Section \ref{NLO}. 

\begin{figure}[h!]
\centering
\includegraphics[scale=1]{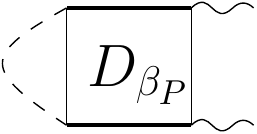}
\caption{Contraction of a type $\beta$ dipole leading to a $2-$point graph. ($P=L$ or $R$)}
\label{dipole_contraction}
\end{figure}

The components that are connected to a broken chain of type $3^*$,$2_L^*$ or $2_R^*$ (on the $\chi$ propagator side) are the NLO schemes described in Section \ref{NLO} with a  $2-$point graph (obtained from the contraction of a dipole of type $\beta_L$ or $\beta_R$ as shown in Figure \ref{dipole_contraction}) inserted in any of their $\chi$ propagator. 
 
\begin{figure}[h!]
\centering
\includegraphics[scale=0.75]{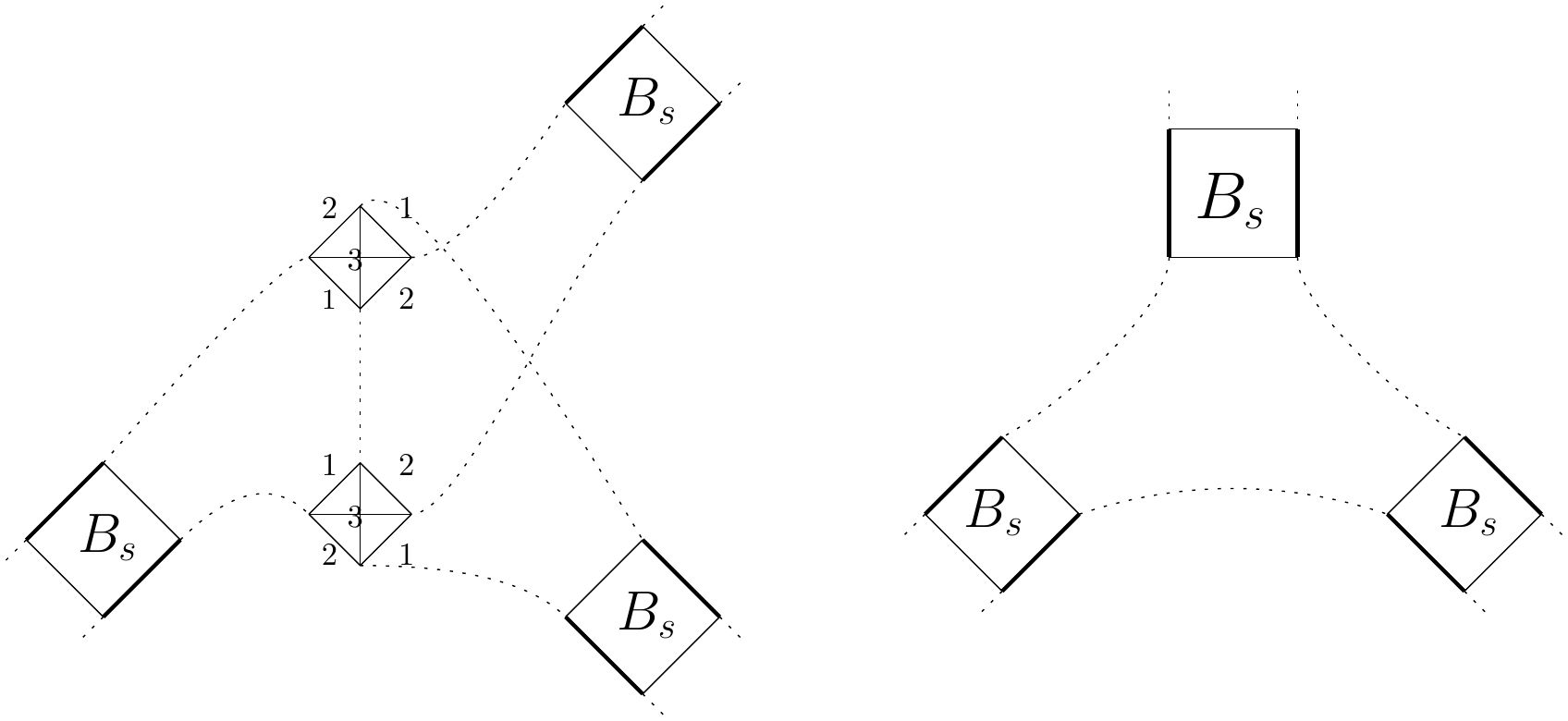}
\caption{The different subschemes that can produce a component of degree 0 and adjacency 3 in the skeleton graph. $B_{t,A}$ can be any broken chain of type $s=1,2_L^*,2_R^*,3^*$ and family A. }
\label{internal_nodes_dominant_scheme}
\end{figure}

The degree $0$ components are all non tracked and have adjacency $3$. They must then come from schemes of the form given by Figure \ref{internal_nodes_dominant_scheme} where $s=1,2_L^*,2_R^*,3^*$. 
%In the skeleton graph $\mathcal{I}'(\mathcal{S}')$, these graphs results in the components given in Figure \ref{all_internal_nodes}.

\begin{figure}[h!]
\centering
\includegraphics[scale=1]{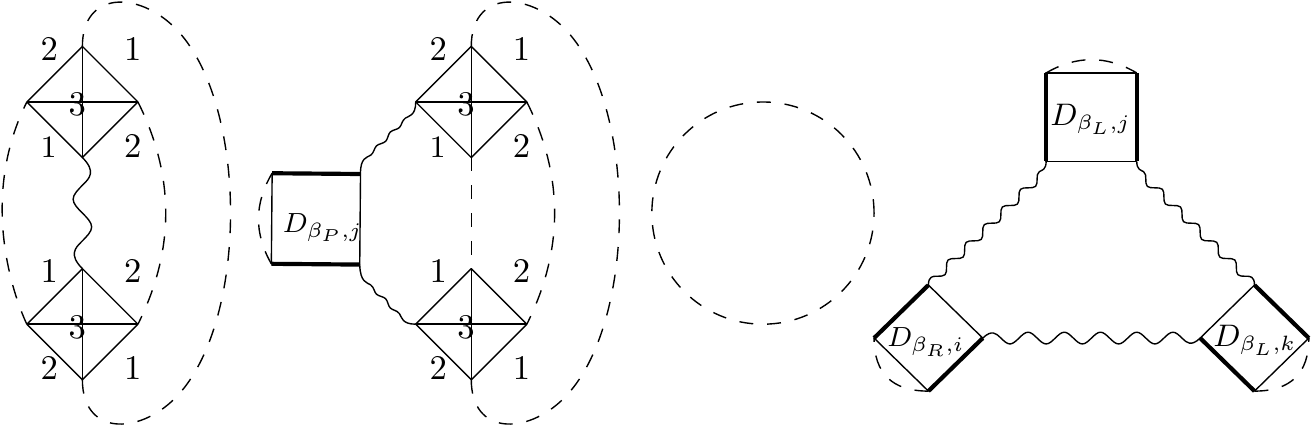}
\caption{The internal nodes of $\mathcal{I}'(\mathcal{S}')$}
\label{all_internal_nodes}
\end{figure} 

%From the adjacency of the different components, 
Finally, 
following \cite{Bonzom_2022}, one gets that the skeleton graph
$\mathcal{I} (\mathcal{S}')$ is a rooted binary tree with $\omega (\mathcal{S})$ leaves given by the rooted component or the degree $1$ components and $\omega (\mathcal{S})-1$ inner nodes given by components of degree $0$.

%one deduces that the degree $0$ components are internal nodes of $\mathcal{I} (\mathcal{S}')$, while the rooted component and the components of positive degree are the leaves of the tree.  

As the schemes are fully encoded by their skeleton graphs, the dominant schemes are thus in bijection with the rooted binary trees described above.
%that are their skeleton graphs $\mathcal{I}' (\mathcal{S}')$.

%\subsection{Generating function of the dominant schemes}

The generating function of these trees is
%in bijection with the dominant schemes of degree $\omega$  is
\begin{equation}
G_{\omega,\mathcal{T}} (t) = B_1(t)^{2 \omega -1} I(t)^{\omega -1} L(t)^{\omega} \text{,}
\end{equation}
where $I(t)$ resp. $L(t)$ are the weights coming from the internal nodes resp. leaves. 

The expression of $L(t)$ is derived by considering all the $2-$point subschemes that lead to a component of degree and valency $1$ in the skeleton graphs $\mathcal{I}'(\mathcal{S})$ of the scheme $\mathcal{S}$. There is a consequent number of such subschemes, hence 
%the consequent number of terms in 
$L(t)$ will be the sum of a consequent number of terms coming from these subschemes.
By a tedious but straightforward computation, one finds 
%that these wieghts are given by 
the following expressions:
\begin{equation}
\begin{split}
I(t) &=  6t M_T(t)^3 M_\chi(t) ( 1+ 3 D_\beta) +1 + (3 D_\beta)^3 \text{,} \\
& =  2(M_T(t) - 1) (1+3 D_\beta) +1 +(3D_\beta^3)
\end{split}
\end{equation}
\begin{equation}
\begin{split}
L(t) &= 9 M_T(t)^9 M_\chi(t)^3 (1+ D_\beta(t)) + 3 C_1(t) + 6 (C_1(t)+ D_\alpha(t)) M^3_T(t) M_\chi(t) \\
& + 12 (C_2(t) + D_\beta(t) )M_T(t)^3 M_\chi(t) + 3 C_5(t) + 6 (C_4(t) + D_\delta(t)) M^3_T(t) M_\chi(t)  \\
&+ 9 D_\beta(t) C_5(t) + 9 D_\beta(t) C_4(t) + 18 C_1(t) M^3_T(t) M_\chi(t) \\
&+ M_T(t)^3 M_\chi(t) \Bigl[18+27 D_\beta(t) + 54 D_\beta (t)+ 54 C_4 (t)+ 18 B_4(t) + 54 C_5(t)  \\
& + 18 B_5(t) \Bigr] + M^3_T (t) M_\chi (t) 27 (1+3 D_\beta (t))\Bigl[ 3 D_\beta^2(t) + 3 D_\beta(t) (C_4(t) + C_5 (t))   \\
& + D_\beta (t) (B_4 (t)+B_5 (t))+ \frac{3}{2} (C_4^2 (t) + C_5^2 (t))+ 3 C_4 (t) C_5 (t) \\
&  + C_4 (t) B_5 (t) + B_4 (t) C_5 (t))\Bigr] +  M^3_T (t)  M_\chi (t) 9 (1+3 D_\beta (t)) (B_4^2 (t) + B_5^2 (t) \\
&  + B_4 (t) B_5 (t)) + 54 M_T (t)^6 M_\chi^2 (t) \Bigl[ 3 D_\beta (t)+ 3 C_4 (t) + B_4 (t) +  3 C_5 (t) \\
&  + 3 D_\beta (t) ( 3 C_4 (t) +  3 D_\beta (t) + 3 C_5 (t) + B_4 (t) + B_5 (t))+ 
  B_5 (t)  + 3 C_4 (t) \Bigl( 3 C_5 (t) \\
& + \frac{3}{2} C_4 (t) + B_4 (t)+ B_5 (t) \Bigr) + 3 C_5 (t) (\frac{3}{2} C_5 (t) + B_4 (t)+ B_5 (t))\Bigr] \text{.}
\end{split}
\end{equation}
%As 
%the generating function 
%$G_{\omega,\mathcal{T}} (t)$ 
%doesn't depend on the shape of the tree, 
We can now sum over all the trees described above and add a melonic insertion at the root to find the generating function of the dominant Feynman graphs  of a given degree $\omega$
\begin{equation}
G_{\omega,\text{dom}} (t) = M_T (t) \sum_{\mathcal{T}} G_{\omega,\mathcal{T}} = M_T (t) \text{Cat}_{\omega -1} B_1(t)^{2 \omega -1} I(t)^{\omega -1} L(t)^{\omega}  \text{,}
\end{equation}
where $\text{Cat}_{\omega -1}$ is the $(\omega-1)^\text{th}$ Catalan number. 
Using equation \eqref{lim_B} we find that

\begin{equation}
G_{\omega,\text{dom}} \stackrel{t \rightarrow t_c}{\sim} 
\text{Cat}_{\omega -1}
M_{T,c} \left(\frac{1}{2 M_{T,c} K \sqrt{1-\frac{t}{t_c}}}\right)^{2 \omega -1} I(t_c)^{\omega -1} L(t_c)^\omega \text{.}
\label{Gdom}
\end{equation}
Let us now define the double scaling parameter $\kappa (t,N)$ as
\begin{equation}
\kappa (t,N) = \frac{I(t_c) L(t_c)}{4 N M_{T,c}^2 K^2 (1-\frac{t}{t_c})} \text{.}
\label{DSL_param}
\end{equation}
Recall that in the $\frac{1}{N}$ expansion, the Feynman amplitude of a generic vacuum graph of degree $\omega$ scales as $N^{3-\omega}$, and 
the one of a $2$-point function graph hence scales as $N^{-\omega}$ (since, when cutting an edge of a vacuum graph in order to obtain a $2-$point function graph, $3$ faces, and hence a factor $N^3$, are lost). 
The contribution of the dominant graphs of degree $\omega$ to the $2$-point function in the double scaling limit then writes 
\begin{equation}
N^{- \omega} G_{\omega,\text{dom}} = M_{T,c} N^{-1/2} \left(\frac{L(t_c)\kappa(t,N)}{I(t_c)}\right)^{1/2} \text{Cat}_{\omega -1} \kappa(t,N)^{\omega -1} \text{.}
\end{equation}

Let us comment on the choice of double scaling parameter. In the expression of 
$N^{-\omega} G_{\omega,dom}$ above, we 
need 
only powers of $\omega$ as exponent of only the factor $\kappa (t,N)$ for two reasons. Firstly, we 
need all non-zero degree dominant graphs to contribute to the same power of $N$ in the double scaling limit. Secondly, by choosing such a double scaling parameter one can sum over all degrees the factor $\text{Cat}_{\omega -1} \kappa(t,N)^{\omega -1}$ to obtain an exact expression for $\sum_{\omega > 0} N^{-\omega} G_{\omega,\text{dom}}$ as given in equation $\eqref{DSL_2_points}$ below.

Adding the contribution of the melons and summing over all the degrees, the 2-point function in the double scaling limit is  
\begin{equation}
\begin{split}
G_{2,DS} (t,N) &= M_{T,c} + \sum_{\omega > 0} N^{-\omega} G_{\omega,\text{dom}} \\
&=  M_{T,c} + M_{T,c} N^{-\frac 12} \left(\frac{L(t_c) \kappa(t,N)}{I(t_c)}\right)^{1/2} \sum_{\omega \in \mathbb{N}^*} \text{Cat}_{\omega -1} \kappa(t,N)^{\omega -1} \\
&= M_{T,c} \left(1 + N^{-\frac 12} \left(\frac{L(t_c)}{I(t_c)}\right)^{1/2} \frac{1-\sqrt{1-4 \kappa(t,N)}}{2 \kappa(t,N)^{1/2}}\right)
\end{split}
\label{DSL_2_points}
\end{equation}

%From equation \eqref{DSL_2_points} 
Let us end this section by giving some interpretation of the result above. 
One can see that, in the double scaling limit, the $2$-point function picks up contributions of all degrees, and not just from the vanishing degree (as it is the case for the large $N$ limit). 
Moreover, one can notice that the higher it is the degree of the graph, the greater it is the contribution 
from the respective degree 
when the coupling constant tends to the critical value.
%The higher the degree of a dominant graph, the more its contribution is important when $t \rightarrow t_c$. 
Note also that, in the limit $\kappa \rightarrow 0$ the large $N$ limit is recovered. This behaviour is identical to the one of the matrix case.

As the sum in equation \eqref{DSL_2_points} is convergent for $\kappa \leq 1/4$, the double scaling limit 
series
of the prismatic model is thus convergent. This is different from the matrix model case where the 
double scaling
series diverge. The result 
obtained here
is analogous to the one obtained for quartic tensor models (see again \cite{Bonzom_2022}, \cite{Gurau_2015} and \cite{Schaeffer}). 
%Therefore, this result is not unique to the quartic case and 
Finally, let us mention that one could expect to obtain a similar double scaling limit behaviour for other sextic tensor models. 

%\begin{enumerate}
 %   \item \`a etoffer la phrase ci-dessous, en insistant que cela est différent du cas de modeles de matrices (classe d'universlite différente)
 %   \item a expliquer que 
 %   plus le degre est grand, plus la contribution est importante quand la constante de couplage tend vers la valeur critique (comme c'est le cas pour les modeles de matrices)
    %contributions from higher degree are enhanced as  $\lambda$ goes to criticality
  %  \item comparaison avec les autres modeles de tenseurs
  %  \item etc.
%\end{enumerate}

\section{Free energy in the double scaling limit}

In this section,
starting from the double scaling limit of the $2-$point function,
we compute the free energy in the double scaling limit .

Recall the relation between the free energy and the $2-$point function 
\begin{equation*}
       F(t,N)  = \tilde F(s=1,\tilde t,N)= - \int_{0}^{1} \frac{ds}{2s} G_2(\tilde t s^3) \text{ .}
\end{equation*}
Since $\tilde F$ is evaluated at $s=1$, taking the limit $t \rightarrow t_c$ is equivalent to taking $\tilde t \rightarrow t_c$.
Recall also that the $2-$point function can be written as
\begin{equation*}
    G_2(t) = M_T(t)+ \sum_{\omega > 0} N^{-\omega} G_{\omega} \text{ .}
\end{equation*}
The free energy then writes:
\begin{equation*}
    F(t,N) = -\int_0^1 \frac{ds}{2s} M_T(s^3 \tilde t) -\sum_{\omega} \int_0^1 
 \frac{ds}{2s}  N^{-\omega} G_{\omega} (s^3 \tilde t) \text{ .}
\end{equation*}
The first term in the equation above is the leading order contribution that we denote by $F_{LO}(t)$. The second term can be computed in the double scaling limit. In this limit only the part of the integral close to one and the dominant graphs contribute to $F(t,N)$. Let us denote by $\epsilon$ a small and fixed quantity. We split the second term as 
\begin{equation*}
  -\sum_{\omega} N^{-\omega} \int^1_{1-\epsilon} 
\frac{ds}{2s} G_{\omega} (s^3 \tilde t) - \sum_{\omega} N^{-\omega}  \int_0^{1-\epsilon} 
  \frac{ds}{2s} G_{\omega} (s^3 \tilde t)  
\end{equation*}
and we are only interested in the first term above
%only keep the part of the integral from $1$ to $1-\epsilon$ as well as 
and in the contribution of the dominant graphs $G_{\omega,dom}$. 
Since %the integral boundaries are close to one and 
the limit $\tilde t \rightarrow t_c$ is taken we can use equation \eqref{Gdom} to rewrite the integral:
\begin{equation*}
    - \sum_{\omega} \int^1_{1-\epsilon} 
 \frac{ds}{2s}  G_{\omega} (s^3 \tilde t) \stackrel{t \rightarrow t_c}{\sim} - \sum_{\omega} \text{Cat}_{\omega -1}
\frac{M_{T,c}  I(t_c)^{\omega -1} L(t_c)^\omega}{(2 M_{T,c} K)^{2\omega -1} }
\int^1_{1-\epsilon} \frac{ds}{2s} (1-\frac{s^3 \tilde t}{t_c})^{1/2- \omega} \text{ .}
\end{equation*}
Using the change of variables $u=s^3$ and $v=1-u \frac{\tilde t}{t_c}$ we find 
\begin{equation*}
    \int^1_{1-\epsilon} \frac{ds}{2s} (1-\frac{s^3 \tilde t}{t_c})^{1/2- \omega} = \frac{3}{2} \int^{1-(1-\epsilon)^3\frac{\tilde t}{t_c}}_{1-\frac{ \tilde t}{t_c}}  dv\frac{v^{1/2-\omega}}{1-v} \text{ .}
\end{equation*}
%Again, since the boundaries of the integral are close to $1-\frac{\tilde t}{t_c}$ 
We can expand $\frac{1}{1-v}$ as 
\begin{equation*}
    \frac{1}{1-v} \approx \sum_{n} (\frac{t_c}{\tilde t})^{n+1}  v^n \text{ .}
\end{equation*} 
The integral is then evaluated as
\begin{equation*}
     \frac{3}{2} \int^{1-(1-\epsilon)^3\frac{\tilde t}{t_c}}_{1-\frac{\tilde t}{t_c}}  dv\frac{v^{1/2-\omega}}{1-v} = \sum_{n} \frac{3}{2 (3/2 - \omega +n)} \left(\frac{t_c}{\tilde t}\right)^{n+1} \left[v^{3/2 - \omega +n}\right]^{1-(1-\epsilon)^3\frac{\tilde t}{t_c}}_{1-\frac{\tilde t}{t_c}} \text{ .}
\end{equation*}
In the limit $\tilde t \rightarrow t_c$, the most divergent contribution is given by the term $n=0$ at $v=1-\frac{\tilde t}{t_c}$. We thus have
\begin{equation*}
    - \sum_{\omega} N^{-\omega} \int^1_{1-\epsilon} 
 \frac{ds}{2s}  G_{\omega} (s^3 \tilde t) \stackrel{\tilde t \rightarrow t_c}{\sim} \sum_{\omega} N^{-\omega}\text{Cat}_{\omega -1}
\frac{M_{T,c}  I(t_c)^{\omega -1} L(t_c)^\omega}{(2 M_{T,c} K)^{2\omega -1} } \frac{\left(1-\frac{t}{t_c} \right)^{3/2 - \omega} }{(1 -2/3 \omega )} \text{ ,}
\end{equation*}
which can be rewritten using the definition of the double scaling parameter \eqref{DSL_param} as: 
\begin{equation*}
    - \sum_{\omega} N^{-\omega} \int^1_{1-\epsilon} 
 \frac{ds}{2s}  G_{\omega} (s^3 \tilde t) \stackrel{\tilde t \rightarrow t_c}{\sim} 
\frac{N^{- 3/2}  I(t_c)^{1/2} L(t_c)^{3/2}}{4 M_{T,c} K^{2} \kappa(t,N)^{1/2} } \sum_{\omega} \text{Cat}_{\omega -1}  \frac{\kappa(t,N)^{\omega-1}}{(1 -2/3 \omega )}   \text{ .}
\end{equation*}
We thus conclude that the free energy in the double scaling limit is given by:
\begin{equation*}
F_{DS}(t,N) = F_{LO,c} + \frac{N^{- 3/2}  I(t_c)^{1/2} L(t_c)^{3/2}}{4 M_{T,c} K^{2} \kappa(t,N)^{1/2} } \sum_{\omega} \text{Cat}_{\omega -1}  \frac{\kappa(t,N)^{\omega-1}}{(1 -\frac 23 \omega )}  \text{ ,}   
\end{equation*}
where $F_{LO,c}$ is the value of $F_{LO}(t)$ at $t= t_c$.

\section{Concluding remarks}

In this paper, we studied the double scaling limit of the $O(N)^3$ invariant-tensor model with a sextic prismatic interaction. Using the intermediate field method, this sextic interaction has been reduced to a quartic $T^3 \chi$ tetrahedric one, $\chi$ beeing the intermediate field. This method allowed us to work in the so-called tetrahedric representation where all prismatic vertices of the Feynman graphs are split in pairs of tetrahedra vertices. In order to obtain our results, we thus generalised the methods used in the study of the single field model with $T^4$ tetrahedric interaction to the $T^3 \chi$ case. 
%From the results obtained in this tetrahedric representation, we finally found the behaviour of the model with the original prismatic interaction by fusing back all the tetrahedric pairs of Feynman vertices into prismatic Feynman vertices.

%In Section \ref{large N}, we have implemented the large $N$ limit of the model and its $\frac{1}{N}$ expansion. To do so, we have derived the scaling parameter value $\alpha$ of the prismatic interaction to be 3 and the expression of the degree of a graph $\omega(\mathcal{G})$.
This strategy allowed us to give a new method to identify the LO graphs explicitly in the large $N$ expansion of the model. In the tetrahedric representation, the LO graphs are melonic graphs with melonic insertions that can be performed on the two types of propagators. In the prismatic representation, the LO graphs are a different class of graphs. These graphs are obtained from the triple tadpole given in Figure \ref{prism_lo} and recursively inserting a $2-$point double tadpole on any propagator or splitting any prismatic vertex into two such vertices linked by a pair of edges as in Figure \ref{prism_insertion}. 
We have then have introduced 
chains, dipoles and schemes 
and we used them to describe the general terms of the $1/N$ expansion of the model. We used them to exhibit both the NLO and the dominant graphs of the model in the tetrahedric representation. The schemes of the dominant graphs are in bijection with rooted binary trees whose leaves are the schemes of the NLO graphs and whose internal nodes are degree $0$ components. Finally, we computed %exactly 
the 2-point function in the double scaling limit, see 
%to be given by 
equation \eqref{DSL_2_points} above. %The results are similar to other tensor models. In the DSL, the 2-point function is convergent for $\kappa \leq 1/4$ and contains the contribution of graphs of any degree. 
 
%Finally, we have identified the 
%schemes of the 
%NLO graphs of the model in the tetrahedric representation and in the prismatic representation. 
%To do so, $\chi$ propagators were inserted in the appropriate schemes.  We then gave the proof of the finitiness of the number of schemes and identified the dominant schemes of the two point function $G(T_1,T_2)$ in Section \ref{Finiteness}. The dominant singularity of $G(T_1,T_2)$ has been found to be at $t_c = \frac{1}{12 \pm 8 \sqrt{2}}$ which is a singular point of both the melons and the broken chains of family A. The dominant schemes, that contribute the most in the limit $t \rightarrow t_c$ maximizes the number of these chains. We have finally found that these schemes are in bijection with rooted binary trees.

\medskip

\begin{figure}[h!]
\centering
\includegraphics[scale=0.7]{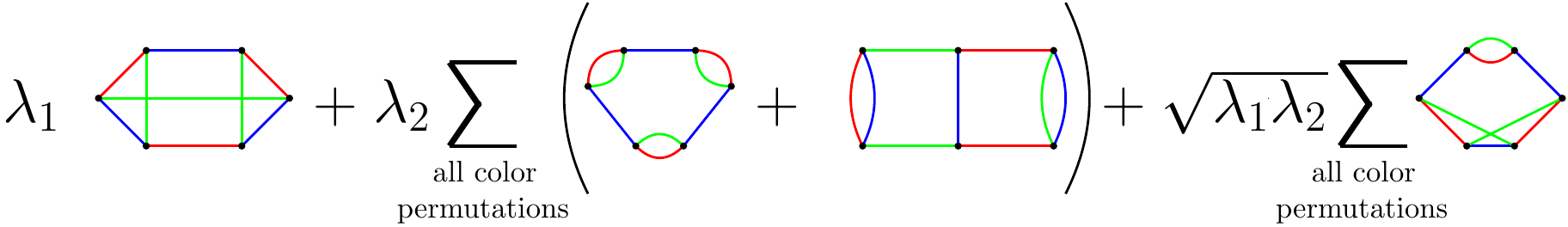}
\caption{The internal nodes of $\mathcal{I}'(\mathcal{S}')$}
\label{sextic_interactions}
\end{figure}

\begin{figure}[h!]
\centering
\includegraphics[scale=0.7]{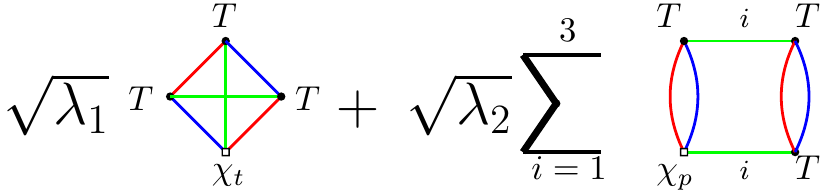}
\caption{The internal nodes of $\mathcal{I}'(\mathcal{S}')$}
\label{quartic_interactions}
\end{figure}

A natural follow up  is to apply  methods 
used in this paper
to other sextic interactions, such as the sextic interactions considered 
(from a renormalization perspective)
in \cite{Giombi_2018}. For example, one could consider a model with the four interactions given in Figure \ref{sextic_interactions}. 
By using a real intermediate field, the interactions of the model can be reduced to a tetrahedric and pillows $T^3 \chi$ interactions as in Figure \ref{quartic_interactions}. In this model, the tetrahedron interactions can be paired to pillow interactions. This renders the model somehow richer than the prismatic model with a more envolved combinatorial structure. %However, we believe that the strategy used in this paper can be applied in order to find the double scaling limit of this model. 
Moreover, the study of this model could serve as an intermediate step toward the computation of the double scaling limit of the $O(N)^3$-invariant tensor model containing all the sextic connected interactions. 

Remark that this model is not the most general as one could consider the same interaction with distinct
coupling constants for each interacting term. In this case, one would need to use both real and complex intermediate fields to express all the sextic interactions in a quartic representation. The presence of  both complex and real intermediate fields 
%might change the analysis presented in this paper.
would thus lead to further complications of the analysis given in this paper.

\begin{figure}[h!]
\centering
\includegraphics[scale=1.5]{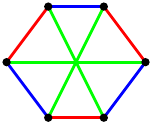}
\caption{The wheel interaction.}% of the prismatic model.}
\label{fig::wheel}
\end{figure}

Moreover, we would like to emphasize that our analysis is not limited to the interactions presented here. In a more general setting, one can defined a separating $D-$cut as a cut of $D$ colored edges that separates a bubble interaction into two disconnected components. Any interaction possessing a separating $D-$cut can be written in an intermediate form using a rank $D$ intermediate field. This intermediate field is real if the two disconnected components are identical and complex otherwise. However, this representation may not be useful as it can lead to interactions  
that cannot be colored in a natural way or that can 
mix tensors of different ranks. When generalising the method described in this paper, one should restrict itself to interactions possessing cuts of the same rank as the one of the tensors in the initial action. 
An example of interaction on which our analysis is not directly applicable is the wheel (or $K(3,3)$) interaction shown in Figure \ref{fig::wheel}. This interaction possesses only separating $4-$cuts (or higher) but is built out of rank $3$ tensors.

\bigskip

{\bf Acknowledgements.}
The authors have been partially supported by the ANR-20-CE48-0018 “3DMaps” grant. A. T. has been partially supported by the PN 09370102 grant. 
The authors acknowledge support of the Institut Henri Poincaré (UAR 839 CNRS-Sorbonne Université), and LabEx CARMIN (ANR-10-LABX-59-01).
The authors warmly acknowledge Victor Nador for  useful discussions at various steps of this research project.

\printbibliography
 
\end{document}